%% file: arxiv_manuscript_gb.tex
\documentclass[conference]{IEEEtran}
\usepackage{fancyhdr}

\IEEEoverridecommandlockouts
\def\BibTeX{{\rm B\kern-.05em{\sc i\kern-.025em b}\kern-.08em
    T\kern-.1667em\lower.7ex\hbox{E}\kern-.125emX}}

\usepackage{comment}
\usepackage{afterpage}
\usepackage{cite}
\usepackage{multirow}
\usepackage{amsmath,amssymb,amsfonts}
\usepackage{algorithmic}
\usepackage{graphicx}
\usepackage{textcomp}
\usepackage[monochrome]{xcolor}
\usepackage{todonotes}
\usepackage{subcaption}
\usepackage{caption}
\usepackage{todonotes}
\usepackage[linesnumbered, ruled]{algorithm2e}
\usepackage{lipsum}
\usepackage{url}
\usepackage{authblk}
\usepackage{blindtext}
\usepackage[switch]{lineno}
\usepackage{booktabs}
\usepackage{mathrsfs} 
\usepackage{listings}
\usepackage{bm}
\usepackage{animate}
\usepackage{algorithmic}
\usepackage{adjustbox}
\usepackage{tabularx}
\usepackage{makecell}

\input{commands.tex}

\begin{document}


\title{Boosting Earth System Model Outputs \\And Saving PetaBytes in their Storage \\ Using Exascale Climate Emulators}



\author[1,6]{Sameh Abdulah}
\author[2,7]{Allison H. Baker}
\author[3,8]{George Bosilca}
\author[4,9]{Qinglei Cao}
\author[5,10]{Stefano Castruccio}
\author[1,6]{Marc G. Genton}
\author[1,6]{\\David E. Keyes}
\author[1,6]{Zubair Khalid}
\author[1,6]{Hatem Ltaief}
\author[1,6]{Yan Song}
\author[1,6]{Georgiy L. Stenchikov}
\author[1,6]{Ying Sun}

\affil[1]{
Extreme Computing \& Statistics \& Earth Science, King Abdullah University of Science and Technology, KSA}
\affil[2]{Computational and Information Sciences Lab, 
NSF National Center for Atmospheric Research, USA}
\affil[3]{NVIDIA, USA} 
\affil[4]{Department of Computer Science, Saint Louis University, USA}
\affil[5]{Department of Applied and Computational Mathematics and Statistics, 
University of Notre Dame, USA}
\affil[6]{\textit {\{Firstname.Lastname\}@kaust.edu.sa}\,\,$^{7}$\textit {abaker@ucar.edu}\,\,$^{8}$\textit {gbosilca@nvidia.com}}
\affil[9]{\textit {qinglei.cao@slu.edu}\,\, $^{10}$\textit {scastruc@nd.edu}}

\maketitle

\let\thefootnote\relax\footnotetext{Authors are listed alphabetically by their last names.}

\thispagestyle{fancy}
\lhead{}
\rhead{}
\chead{}
\renewcommand{\headrulewidth}{0pt} \renewcommand{\footrulewidth}{0pt}

\setboolean{GuideOn}{false}

\begin{abstract}
\input{abstract.tex}
\end{abstract}

\vspace{0.3in}

\begin{IEEEkeywords}
Dynamic runtime systems, High-performance computing, Mixed-precision computation, Spatio-temporal climate emulation, Spherical harmonic transform, Task-based programming models.
\end{IEEEkeywords}




\section{Introduction}
\input{overview.tex}
\section{CURRENT STATE OF THE ART}
\input{soa.tex}

\section{INNOVATIONS REALIZED}
\input{innovations.tex}

\section{HOW PERFORMANCE WAS MEASURED}
\input{measures.tex}

\section{PERFORMANCE RESULTS}
\input{performance.tex}

\section{IMPLICATIONS}
\input{implications.tex}

\section*{ACKNOWLEDGMENTS}
\input{ack.tex}



\bibliographystyle{unsrt2authabbrvpp}
\bibliography{arxiv_manuscript_gb}

\end{document}

%% file: commands.tex
\usepackage{xspace}
\usepackage[nolist]{acronym}
\usepackage{enumerate}
\usepackage{enumitem}
\usepackage{ifthen}

\newboolean{GuideOn} 
\setboolean{GuideOn}{True}

\newcommand{\frontierNode}{{$9{,}025$}\xspace}
\newcommand{\frontierPerf}{{$0.976$}\xspace}
\newcommand{\frontierGPU}{{{$36{,}100$}}\xspace}

\newcommand{\datapointsinbillion}{{{$318$}}\xspace}
\newcommand{\datapointsinbilliondaily}{{{$31$}}\xspace}

\newcommand{\bandlimits}{{$720${,} $1{,}440${,} $2{,}880$, and $5{,}219$}\xspace}  

\newcommand{\ALPSNode}{{$1{,}936$}\xspace}
\newcommand{\ALPSPerf}{{$0.739$}\xspace}
\newcommand{\ALPSGPU}{{$7{,}744$}\xspace}

\newcommand{\leonardoNode}{{$1{,}024$}\xspace}
\newcommand{\leonardoPerf}{{$0.243$}\xspace}
\newcommand{\leonardoGPU}{{$4{,}096$}\xspace}

\newcommand{\summitNode}{{$3{,}072$}\xspace}
\newcommand{\summitPerf}{{$0.375$}\xspace}
\newcommand{\summitGPU}{{$18{,}432$}\xspace}

\newcommand{\resolutionDegree}{ $0.034^\circ$\xspace}
\newcommand{\resolutionKm}{$3.5$ km\xspace}

\newcommand{\densemp}{\texttt{MP+dense}\xspace}
\newcommand{\densetlrmp}{\texttt{MP+Dense/TLR}\xspace}

\newcommand{\fugaku}{\texttt{Fugaku}\xspace}

\newcommand{\parsec}{\texttt{PaRSEC}\xspace}
\newcommand{\starpu}{\texttt{StarPU}\xspace}
\newcommand{\charm}{\texttt{Charm++}\xspace}

\newcommand{\ompss}{\texttt{OmpSs}\xspace}

\newcommand{\hpx}{\texttt{HPX}\xspace}
\newcommand{\legion}{\texttt{Legion}\xspace}

\newcommand{\reduce}[1]{}

%% file: abstract.tex
\ifthenelse {\boolean{GuideOn}} {\textbf{(150 word max)}\\}{}

We present the design and scalable implementation of an exascale climate emulator for addressing the escalating computational and storage requirements of high-resolution Earth System Model simulations. We utilize the spherical harmonic transform to stochastically model spatio-temporal variations in climate data. This provides tunable spatio-temporal resolution and significantly improves the fidelity and granularity of climate emulation, achieving an ultra-high spatial resolution of \resolutionDegree ($\sim$\resolutionKm) in space. Our emulator, trained on \datapointsinbillion billion hourly temperature data points from a 35-year and \datapointsinbilliondaily billion daily data points from an 83-year global simulation ensemble, generates statistically consistent climate emulations. We extend linear solver software to mixed-precision arithmetic GPUs, applying different precisions within a single solver to adapt to different correlation strengths. The \parsec runtime system supports efficient parallel matrix operations by optimizing the dynamic balance between computation, communication, and memory requirements. Our BLAS3-rich code is optimized for systems equipped with four different families and generations of GPUs, scaling well to achieve \frontierPerf EFlop/s on \frontierNode nodes (\frontierGPU AMD MI250X multi-chip module (MCM) GPUs) of Frontier  (nearly full system), \ALPSPerf EFlop/s on \ALPSNode nodes (\ALPSGPU Grace-Hopper Superchips (GH200)) of Alps, \leonardoPerf EFlop/s on \leonardoNode nodes (\leonardoGPU A100 GPUs) of Leonardo, and \summitPerf EFlop/s on \summitNode nodes (\summitGPU V100 GPUs) of Summit.

%% file: overview.tex
\ifthenelse {\boolean{GuideOn}} {
{\bf description of the problem and its importance, in terms understandable to a non-specialist (1 p max)}
} {
}

Climate change, evident in rising temperatures, extreme weather events, sea-level rise, and ecosystem disruption, poses significant risks and urgently requires action due to intensified heatwaves, storms, droughts, floods, and biodiversity loss~\cite{Rising:2022,IPCC:AR6:2022}. We stand at a critical juncture where converging advancements in technology and methodologies, including the integration of exascale supercomputers, offer unprecedented opportunities to tackle the existential threats of climate change. Our work leverages exascale computing to develop scalable climate emulators that complement high-resolution climate projections, a capability not practically affordable by traditional climate models alone due to their expensive computational demands and petabyte-scale storage requirements.

Conventional climate models use conservation laws expressed as partial differential equations (PDEs) to represent the interactions within Earth's systems, providing insights into climate change causes, mitigation impacts, and the potential trajectories of future climate conditions~\cite{Sognnaes:2021,Mort:2007}. Climate modeling began in the 1950s, and over the years, models have become increasingly sophisticated as computing power, observations, and our understanding of the climate system have advanced, with Earth System Models (ESMs) now  the state-of-the-art models available to climate scientists~\cite{Flato:2014}.

Throughout 80 years of modern computing, capabilities deemed unimaginable by one generation have become routine expectations of the next. Gordon Bell papers have regularly expanded the headroom of applications, in terms of resolution, fidelity, and/or performance \cite{DongarraKeyes2024}, which we do here for the next Coupled Model Intercomparison Project (CMIP) campaign. The CMIP campaigns support detailed comparisons of ESMs by enabling systematic analysis, validation, and enhancement. CMIPs are ``resource-intensive and time-consuming to run"\cite{mickelson2020}, with each new phase becoming increasingly complex in terms of experimental design, number of model intercomparisons, and the growth in data requested.
For example, the most recent CMIP Phase 6 (CMIP6) effort resulted in roughly 28 PB from 45 modeling institutes hosted by the Earth System Grid Federation~(ESGF)~\cite{ESGF}.
For comparison, CMIP3 generated 40 TB of data, and CMIP5 generated 2 PB of data~\cite{cmip6}. For individual modeling centers, generating and preparing the data for the CMIP efforts is typically a monumental task. 
For CMIP6, the National Center for Atmospheric Research (NCAR) ran 37,000 years of climate with the Community Earth System Model (CESM), consuming 190 million CPU hours and resulting in 2 PB post-processed time series data~\cite{mickelson2020}. Similarly, the Goddard Institute for Space Studies (GISS) contribution to CMIP6 required 104.5 million CPU hours on the NASA Center for Climate Simulation (NCCS) Discover supercomputer and generated 147 TB of simulation data~\cite{nasa}. Moreover, financial constraints limit the storage capacity available for climate simulations. For example, managing data at NCAR incurs costs of approximately \$45 per TB annually. This results in substantial financial burdens for projects with petabyte-scale storage needs and can limit science objectives. In a recent survey of the NCAR computing community, 70\% of
respondents indicated that disk volume limitations have a moderate or
severe impact on their work.

Computational demands and storage requirements for ESMs continue to escalate as the climate community progresses toward ultra-high-resolution (i.e., kilometer-scale) simulations. Some scientists believe that simulations at these so-called ``global storm-resolving" scales are necessary to understand better how weather and extremes will be affected by climate change~\cite{KilometerScaleClimateModelsProspectsandChallenges}.
The recent DYAMOND (DYnamics of the Atmospheric general circulation Modeled On Non-hydrostatic Domains) project~\cite{Dyamond} exemplifies the computational cost of kilometer-scale simulations. In the initial DYAMOND experiments, the nine atmosphere-only models in the project ran for only 40 days, due to both computational and petabyte-scale storage costs (note that short timesteps are typically needed for high-res model runs). 
At the grid spacing of 5 km or less needed to capture small-scale processes, the average DYAMOND model output rate was just 6 SDPD (simulated-days-per-day), and 3D variable output was severely limited.  For example,  each output sample for the ICON model simulations was $\sim$1 TB~\cite{Dyamond}, and
the 3.25 km resolution SCREAM (Simple Cloud-Resolving E3SM Model, whose developers were awarded the first Gordon Bell Prize for Climate Modeling in 2023~\cite{e3sm-scream2023}) contribution to DYAMOND produced $\sim$4.5 TB of data per simulated day~\cite{caldwell2021}.
Note that even a coarser quarter-degree horizontal-grid resolution ($\sim$25 km), which is still considered high-resolution for climate, is quite costly.  For example,  
E3SM consumed almost 100 million core hours on Argonne's Theta supercomputer for a single 50-year-long simulation~\cite{Caldwell:2019}. 
Because the model resolution significantly impacts the trade-off between the computational costs and representation of the climate system, addressing both computational and data storage challenges is essential for advancing climate modeling capabilities.
\begin{figure*}[!ht]
        \centering
                \includegraphics[width=0.63\textwidth]{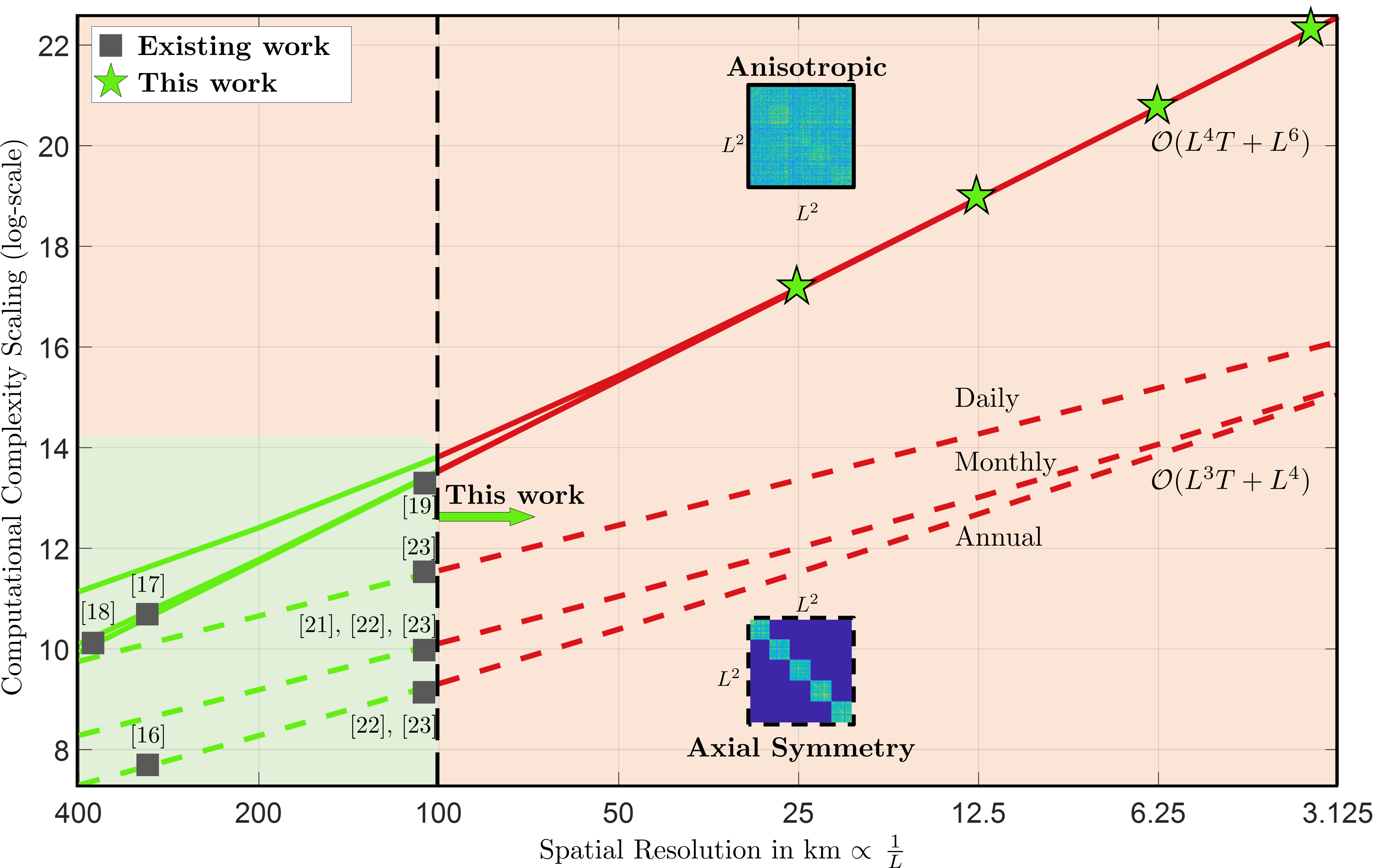}
        \caption{\small{Overview of existing work vs. this work. Analysis of computational cost associated with the emulator design for different spatial resolutions and the review of the existing emulators. The computational complexity scales as $ \mathcal{O}(L^3T+L^4)$ and $\mathcal{O}(L^4T+L^6)$ for axially symmetric and longitudinally anisotropic models, respectively, where $T$ is the count of temporal data points and $L$ parameterizes the spatial resolution, reflecting the number of grid points along either latitude or longitude. Existing emulators have achieved spatial resolutions of up to 100 km (approximately 0.901 degrees at the equator) and temporal resolutions down to the daily scale under the axial symmetry~(stationarity with respect to longitude) assumption. In contrast, anisotropic models have been limited to an annual temporal scale at 100 km spatial resolution. The climate emulator design in this work (indicated by a green star) significantly advances the fidelity and granularity of climate emulation by achieving an unprecedented spatial resolution ($\sim$\,3.5--100 km)} and hourly temporal resolution, surpassing the existing emulators by a factor of 245,280 (28 times in spatial resolution and 8,760 times in temporal resolution). 
}
    \label{fig_overview}
    \vspace{-4mm}
\end{figure*}

\begin{figure}[t]
    \begin{minipage}[b]{0.5\textwidth}
        \centering
        \begin{subfigure}[b]{0.5\textwidth}
            \centering
            \animategraphics[loop,trim = 4cm 1cm 3.5cm 1cm,autoplay,scale=0.5,controls=play]{3}{figures_new/animations/frame_jan_}{1}{12}
            \caption{\small{Jan. 01, 2019 Simulations}}
        \end{subfigure}%
        \hspace{-2mm}
        \begin{subfigure}[b]{0.5\textwidth}
            \centering
            \animategraphics[loop,trim = 4cm 1cm 3.5cm 1cm,autoplay,scale=0.5,controls=play]{3}{figures_new/animations/frame_e35_jan_}{1}{12}
            \caption{\small{Jan. 01, 2019 Emulations}}
        \end{subfigure} \\[2mm]
        \begin{subfigure}[b]{0.5\textwidth}
            \centering
            \animategraphics[loop,trim = 4cm 1cm 3.5cm 1cm,autoplay,scale=0.5,controls=play]{3}{figures_new/animations/frame_jun_}{1}{12}
            \caption{\small{Jun. 01, 2019 Simulations}}
        \end{subfigure}%
        \hspace{-2mm}
        \begin{subfigure}[b]{0.5\textwidth}
            \centering
            \animategraphics[loop,trim = 4cm 1cm 3.5cm 1cm,autoplay,scale=0.5,controls=play]{3}{figures_new/animations/frame_e35_jun_}{1}{12}
               \caption{\small{Jun. 01, 2019 Emulations}}
     \end{subfigure}
    \end{minipage}

    \begin{minipage}[b]{0.5\textwidth}
    \centering
        {\includegraphics[width=0.9\textwidth]{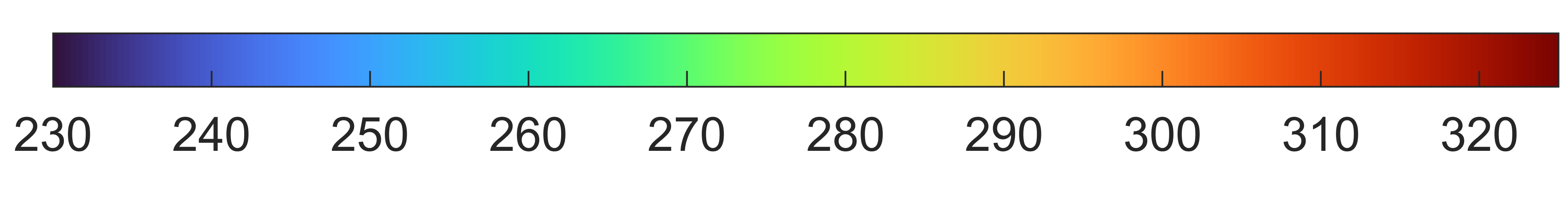}}
      \end{minipage}
    \caption{\small{Hourly (24-hour) surface temperatures (in K) generated by climate simulations (ERA5 dataset) and the proposed climate emulator (single realization, statistically consistent with simulations) are plotted in (a) simulations and (b) emulations for January 1, 2019, and in (c) simulations and (d) emulations for June 1, 2019. This figure includes an animation (at 2-hour intervals) for enhanced visualization, optimally viewed in {\it Adobe~Acrobat}.}}
    \vspace{-4mm}
    \label{fig_illustraion}
\end{figure}

Climate emulators play a pivotal role in alleviating the computational burden and storage requirements associated with climate modeling and simulations, as highlighted by their use in the Intergovernmental Panel on Climate Change (IPCC) assessment report (AR6) to forecast warming across various emission scenarios. These emulators, capable of controlling numerous climate system aspects through parameter variation, can be tuned to mimic ESMs behavior using simulated data such as CMIP6 and generate climate projections within seconds on a standard desktop computer. This enables multiple runs with varied parameter values for a single emissions scenario, which is crucial for encompassing the uncertainty range in future climate projections. Emulators have been part of the climate modeling landscape for almost as long as the development of more advanced climate models~(e.g., \cite{Castruccio:2013,Castruccio:2014,Holden:2015,Link:2019,Castruccio:2019,Jeong:2019,Huang:2023,Song:2023}). In Figure~\ref{fig_overview}, we review these contributions through the lens of the computational cost associated with climate emulator design~(more details provided in the next section). We note that the current climate emulators have not yet attained a spatial resolution finer than 100 km. Moreover, only those emulators designed with an assumption of axial symmetry~(stationarity with respect to longitude) have managed to achieve a daily temporal resolution. In this context, our work makes the following notable contributions:

\begin{itemize}
\item 
We have developed a computationally efficient exascale climate emulator that surpasses existing emulators by a factor of 245,280 in terms of spatio-temporal resolution, that is, offering 28 times greater spatial and 8,760 times greater temporal resolution (see Figure~\ref{fig_overview}). This advancement marks a significant milestone in climate science, as it enables the generation of hourly emulations at an ultra-high spatial resolution of \resolutionDegree ($\sim$\resolutionKm) for the first time --- a step towards accurate climate projections and a deeper understanding of climate change and its impacts. Our emulator, capable of generating climate emulations statistically consistent with the simulations, is trained on \datapointsinbillion billion hourly and \datapointsinbilliondaily billion daily data points of surface temperature derived from a global ensemble of simulations spanning 35 years (1988-2022) and 83 years (1940-2022), respectively.
\item Our approach addresses the limitations of existing emulators by employing the spherical harmonic transform (SHT) of global data to model anisotropic interactions. We also present HPC innovations for the scalable implementation of the emulator design on the exascale computers. Our implementation applies multiple precisions within a single solver, catering to the covariance strengths of the data. Furthermore, we rely on \parsec runtime system to support efficient parallel matrix operations and optimize the balance between computation, communication, and memory requirements with tile-centric mixed-precision support. The scalability of the design is demonstrated by generating unprecedented \resolutionDegree ($\sim$\resolutionKm) hourly climate emulations, equivalent to $477$ billion data points for a single year emulation of temperature data, illustrated in Figure~\ref{fig_illustraion} at $25$ km spatial resolution (the maximum available resolution of ERA5 data). 
\item We optimize our code for large-scale execution on systems equipped with different families and generations of GPUs, scaling well to achieve \frontierPerf EFlop/s on \frontierNode nodes (\frontierGPU AMD MI250X GPUs) of Frontier (topping the June 2024 Top500 list), \ALPSPerf EFlop/s on \ALPSNode nodes (\ALPSGPU NVIDIA H100 GPUs) of Alps (ranked $6^{th}$), \leonardoPerf EFlop/s on \leonardoNode nodes (\leonardoGPU NVIDIA A100 GPUs) of Leonardo (ranked $7^{th}$), and \summitPerf EFlop/s on \summitNode nodes (\summitGPU NVIDIA V100 GPUs) of Summit (ranked $9^{th}$). Our implementation demonstrates excellent weak scaling efficiency and up to 72\% strong scaling efficiency with as many as $12{,}288$ V100 GPUs on Summit. We also rely on KAUST's Shaheen III CPU partition (ranked $23^{rd}$) to perform most of the science experiments in this paper.
\end{itemize}


We thus bridge the gap between ESM simulations and the growing demand for accessible, highly resolved emulations and projections. Since our emulator enhances and complements computationally expensive and storage-intensive simulations, we anticipate substantial savings in storage (in petabytes) needed for storing such simulations, heralding a new era of analysis at ultra-high spatio-temporal resolutions. Furthermore, by shifting the majority of the computational effort from communication-bound sparse double-precision kernels characteristic of PDE-based models to dense low-precision tensor core kernels in the wheelhouse of the GPU commodity market, we open a more sustainable swim lane to climate modeling.

%% file: soa.tex
\ifthenelse {\boolean{GuideOn}} {
\textbf{quantitative discussion of current SoA, with emphasis on performance-related aspects  (1 p max)\\
}
} {
}
\begin{figure*}[!h]
\center
\vspace{-2mm}
\hspace{-2mm}
\includegraphics[width=0.9\linewidth]{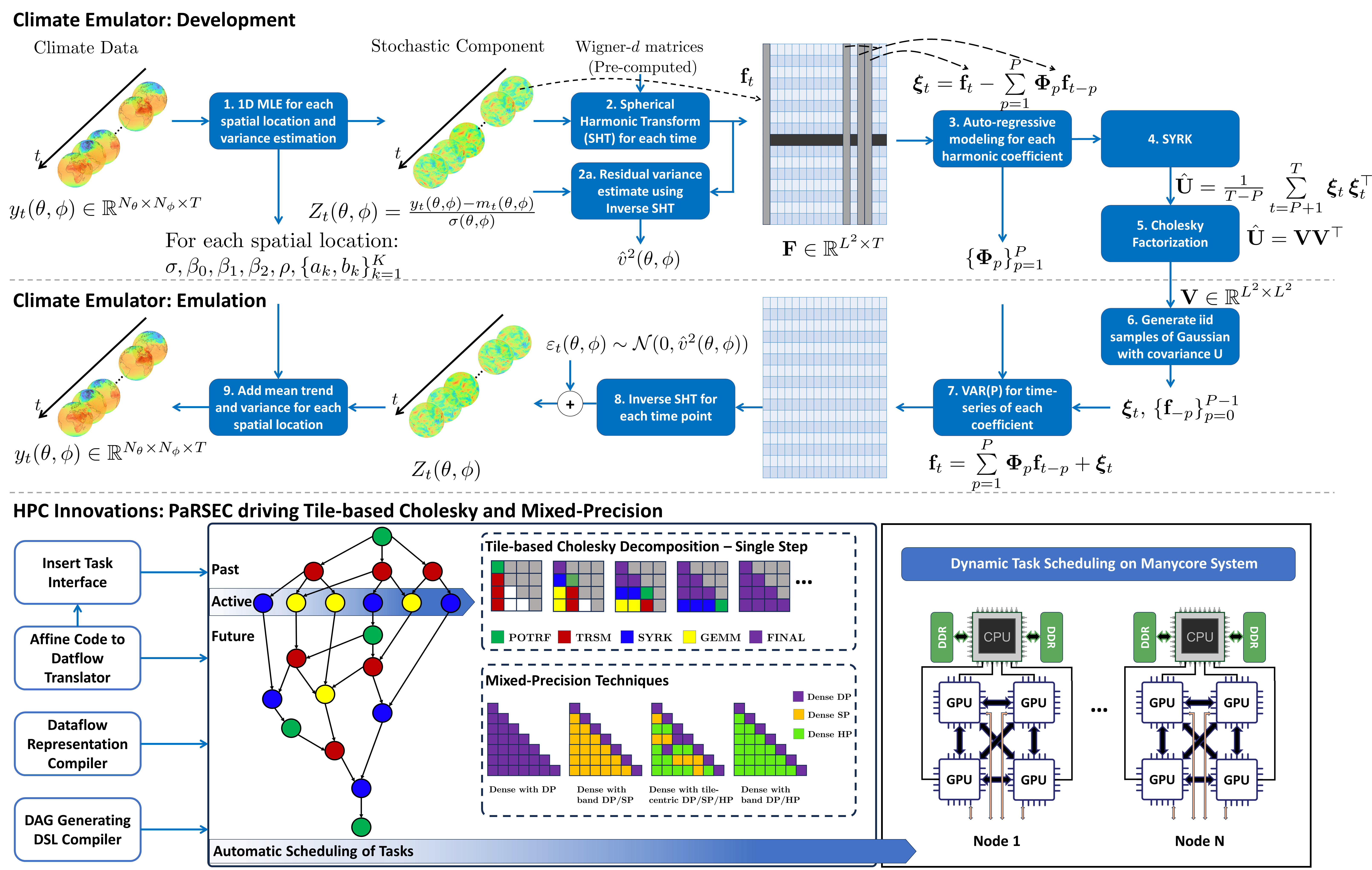}
\caption{{\small
An overview of the design and development pipeline of the climate emulator, including HPC innovations, detailing each step in the process of developing an emulator and using it to generate emulations~(see Section \ref{sec:innovations} for details).}}
\vspace{-4mm}
\label{fig:emulator_design}
\end{figure*}


\subsection{Climate Emulators:  Review and Computational Challenges}

The development of climate emulators has recently gained prominence due to their role in complementing and enhancing the ESM simulations~(e.g.,\cite{Oakley2004Probabilistic,chang2014,Castruccio:2014,Hu2021Approximating,Branstator:2010}, see Section \ref{sec:implications} for potential uses of climate emulators). At the heart of emulator design is the principle of training a statistical model on a limited set of simulations to generate new data or predict outcomes for new inputs, eliminating the need for a large ensemble of full-scale climate model runs. The development of emulators spans decades, with the conventional approach primarily revolving around the use of Gaussian processes~(GPs) and auto-regressive models to capture the intricate spatio-temporal dependencies found in climate simulations driven by underlying physical mechanisms. A primary challenge with this approach is the processing of extensive high-resolution data in global climate simulation, as performing inference with GPs on such large datasets can be computationally prohibitive. Addressing this computational challenge benefits from drawing on existing techniques such as low-rank approximations~\cite{FRK2008}, composite likelihoods~\cite{guinness2019gaussian}, stochastic partial differential equations~\cite{bolin2023covariancebased}, covariance tapering \cite{taper2008}, and their combinations~\cite{datta2016hierarchical}. Notably, efficient algorithms to handle large datasets and the distribution of computations across multiple processors have been proposed to reduce the emulation times~\cite{Castruccio:2014,Castruccio:2013}. \cite{Song:2023} presents an innovative emulator tailored for global surface temperature simulations from the recently published CESM2-LENS2 data, featuring a spatial resolution of approximately $1^\circ \times 1^\circ$. By leveraging the SHT, the emulator efficiently transitions simulations from the spatial domain to the spectral domain. Exploiting continuity by changing the discrete basis from grid values to spectral amplitudes leads to significant reductions in both the computational and storage costs, enabling the generation of emulations across annual, monthly, and daily scales. The spectral basis also provides a unified representation of data with different grid resolutions.

The anisotropic (non-stationary) spatial dependence in the output of ESM simulations also presents significant computational challenges associated with estimating and decomposing the covariance matrix that captures this spatial dependence. Common simplifications include assuming isotropy and axial symmetry~(stationarity along the longitude), leading to diagonal and sparse covariance matrices, respectively~\cite{Castruccio:2013,Jeong:2019,Huang:2023, Song:2023}. This reduces the computational workload for matrix estimation and its Cholesky decomposition. 

In Figure~\ref{fig_overview}, we analyze and compare the spatial resolution and computational cost of various emulators proposed in the literature. The computational complexity when assuming axial symmetry scales as $\mathcal{O}(L^3T+L^4)$, where $T$ denotes the number of temporal data points, and $L$ denotes the spatial resolution, reflecting the number of grid points along either latitude or longitude. Emulators designed under this assumption have reached spatial resolutions up to 100 km and temporal resolutions down to the daily scale.  Relaxing the axial symmetry assumption increases the computational complexity associated with the emulator design to $\mathcal{O}(L^4T+L^6)$. As a result, emulators accommodating longitudinal variation are typically confined to annual temporal resolutions and moderate spatial resolutions ($\sim$100-500 km)~\cite{Holden:2015,Castruccio:2014,Link:2019}. We overcome the computational challenges associated with accurately capturing the anisotropic spatial dependencies in ESMs by employing HPC innovations in our scalable implementation. These innovations enable efficient parallel implementation of SHT and the computation and decomposition of the covariance matrix. The proposed HPC innovations, combined with recent scientific advancements, offer pathways to harness the power of supercomputers to develop exascale scalable climate emulators capable of generating climate emulations consistent with the intricate details of climate dynamics and achieving a level of resolution previously unattainable in both space and time. 

\subsection{Climate Emulator Design -- Overview}

We consider spatio-temporal climate data, denoted by $y_t^{(r)}(\theta_i,\phi_j)$ for latitude $\theta_i\in[0,\pi]$, longitude $\phi_j\in[0,2\pi)$, time point $t$, and ensemble $r$, where $i=1,\ldots,N_\theta$, $j=1,\ldots,N_\phi$, $t=1,\ldots,T$, and $r=1,\ldots,R$. The data in the spatial domain are located on a grid of $N_\theta \times N_\phi$ points, and the total size of the dataset given by $R \times T \times N_\theta \times N_\phi$ depends on the spatial and temporal resolutions. 
As an illustration, hourly observations of ERA5 surface temperature are plotted in Figure \ref{fig_illustraion}(a) and (c).

Utilizing the SHT, which transforms spatial data on the sphere into the spectral domain based on eigenfunctions of the angular part of the 3D Laplacian, we present the design of an emulator equipped with efficient and adaptable capabilities tailored specifically for managing the complexities inherent in large-scale global climate simulations such as non-stationary spatial dependencies that exist among latitudes and between different geographic regions~(e.g., land and ocean), and temporal dynamics. The capability of the proposed emulator for generating climate emulations {\it statistically consistent} with the simulations has been demonstrated in \cite{Song:2023} for different scenarios. Figure \ref{fig:emulator_design} provides an overview of the emulator design and the steps involved in generating emulations, including HPC innovations. In the next section, we describe the mathematical formulation and computational complexity associated with designing the emulator and generating emulations.

\subsection{Advancements in Dense Linear Algebra: State-of-the-Art}
Dense matrix algorithms are extensively utilized across various applications, from scientific computing to engineering and data analysis. Operations such as Cholesky have seen significant performance improvements due to rapid advancements in hardware technology. With well-designed parallel algorithms, the performance of these dense operations can approach the underlying theoretical limits of the hardware. 
These algorithms use task-based programming to handle detailed computations efficiently and are known as tile-based linear solvers. They divide the matrix into tiles represented in a Directed Acyclic Graph (DAG) in Figure ~\ref{fig:emulator_design}, where nodes and edges represent computational tasks and their dependencies, respectively~\cite{agullo2009numerical}.

On modern hardware, such as GPUs, there is a trend towards supporting operations at lower precision, a development that AI workloads capitalize on for accelerated training and inference. For instance, NVIDIA V100, A100, and H100  can run single-precision (SP) / half-precision (HP) 2X/16X,  16X/32X, 14.7X/29.5X faster than double-precision (DP) arithmetic, respectively (assuming, if supported, tensor core operations and the SXM interface). This shift also presents an opportunity in dense linear algebra to exploit fast matrix operations performed at reduced precision, particularly in scientific applications that can tolerate reduced computational accuracy.

Mixed-precision matrix approximations capitalize on hardware trends and substantially reduce memory usage. This reduction allows more of the working set to reside in fast memory, enhancing performance \cite{abdulah2021accelerating, cao2023reducing}. To effectively manage these operations, which vary inherently in their data structures and precision levels (i.e., DP/SP/HP), it is essential to implement a dynamic runtime system framework that adeptly orchestrates these heterogeneous tasks.

\subsection{Increased Productivity with Dynamic Runtime Systems}

Task-based dynamic runtimes have been proposed as an alternative for programming challenges raised by complex hierarchical systems, with the goal of facilitating the scheduling of fine-grained ``work units", or tasks, onto the underlying hardware resources and managing the communications. Augmented with various scheduling strategies, they are able to reduce idle time for imbalanced applications and reduce expensive data movements while favoring data locality. These runtime systems support shared- and distributed- memory systems, including support for accelerators. 
The result is an asynchronous execution model offering adaptive capabilities beyond the bulk synchronous SPMD mainstream. 

The programming API varies between the different frameworks, going from a relatively easy to depict sequential declaration of tasks and their dependencies for OpenMP~\cite{OpenMP}, \starpu~\cite{augonnet2011starpu}, \ompss~\cite{ompss}, \hpx~\cite{Heller2013a}, \charm~\cite{kale.oopsla.1993}, to a more complex, static or dynamic, symbolic representation of the graph of tasks for \parsec~\cite{bosilca2013parsec} and \legion~\cite{Bauer2012}. In all cases, the users remain responsible for the correct description of the graph of tasks, and the runtime is responsible for the effective unrolling of the tasks and resulting communications onto the available resources, providing a separation of concerns between algorithmic description and execution.

\parsec~\cite{bosilca2013parsec}, the runtime used in this work, relies on a domain-specific language (DSL) that represents the entire DAG in a compressed and parametrized manner. This permits leveraging collective communications that are crucial in dense matrix algorithms, provides opportunities for communication/computations overlap, and better uses data locality and caching. \parsec decouples the data distribution from the actual task operations, providing a level of abstraction allowing for type changes, or adaptive precision conversion, at the communication level, based on the needs for successors.

%% file: innovations.tex
\ifthenelse {\boolean{GuideOn}} {
\textbf{what the innovations are and how they were achieved (2 pp max)\\
}
} {
}
\label{sec:innovations}




\subsection{Climate Emulator Design -- Formulation}
 
We represent the climate data in terms of deterministic and stochastic components as~\cite{Huang:2023,Song:2023}
\begin{equation}
    y_t^{(r)}(\theta_i,\phi_j) = m_t(\theta_i,\phi_j)+\sigma(\theta_i,\phi_j)Z_t^{(r)}(\theta_i,\phi_j),
    \label{eq:decouple}
\end{equation}
where $Z_t^{(r)}(\theta_i,\phi_j)$ is the stochastic component at spatial location $(\theta_i,\phi_j)$,  time point $t$, and ensemble $r$. The terms $m_t$ and $\sigma$ represent the deterministic functions responsible for the mean trend and standard error, respectively, and are shared across all ensembles. The mean trend $m_t$ for each spatial location can be modeled using an infinite distributed lag model to relate it with the radiative forcing (RF) trajectory $x_t$~(annual scale) as~\cite{Castruccio:2014,Huang:2023,Song:2023}
\begin{align}
m_t=&\beta_0+\beta_1 x_{\lceil t/\tau \rceil}
 +\beta_2(1-\rho)\sum_{s=1}^{\infty}\rho^{s-1}x_{\lceil t/\tau \rceil-s},
\nonumber \\
&+\sum_{k=1}^{K}\left\{a_k\cos\left(\frac{2\pi tk}{\tau}\right)+b_k\sin\left(\frac{2\pi tk}{\tau}\right)\right\},
\label{Eq:mt_model}
\end{align}
where $\beta_0(\theta_i,\phi_j)$ is the intercept, and  $\beta_1(\theta_i,\phi_j)$ and $\beta_2(\theta_i,\phi_j)$ are slopes for the current and past RF, respectively. Here $\beta_2(\theta_i,\phi_j)\{1-\rho(\theta_i,\phi_j)\}\rho(\theta_i,\phi_j)^{s-1}$ are the lag weights that decrease the impact of past RF exponentially by $\rho(\theta_i,\phi_j)\in[0,1]$. The choice of $K$ depends on the number of terms used to model the periodic (e.g., intraday or interannual) variations. $\tau$ is a parameter capturing the frequency of periodic variations with a value $\tau=12$, $\tau=365$ and $\tau=8760$ for monthly, daily, and hourly temporal resolutions, respectively. Leveraging the independence, we can efficiently estimate the parameters in \eqref{Eq:mt_model} and $\sigma$ by using 1D maximum likelihood estimation (MLE) method~\cite{Huang:2023,Song:2023} for each spatial location with the computational complexity $\mathcal{O}(T)$. 

\subsubsection{Modeling in the Spherical Harmonic Domain}
We model the stochastic component in terms of spherical harmonics as
\begin{equation*}
    Z_t^{(r)}(\theta_i,\phi_j)=\sum_{\ell=0}^{L-1}\sum_{m=-\ell}^\ell (z_t^{(r)})_{\ell,m} Y_{\ell,m}(\theta_i,\phi_j)+\varepsilon_t^{(r)}(\theta_i,\phi_j),
\end{equation*}
where $Y_{\ell,m}(\theta_i,\phi_j) = \sqrt{\frac{2\ell+1}{4\pi}\frac{(\ell-m)!}{(\ell+m)!}}P_\ell^m(\cos\theta_i)\exp(\mathrm{i} m \phi_j)$ and $(z_t^{(r)})_{\ell,m}$ represent an orthonormal spherical harmonic basis function and coefficient, respectively, of degree $\ell\ge0$ and order $|m|\le \ell$. Since we have truncated the representation in terms of basis functions at degree $L-1$, the last term $\varepsilon_t^{(r)}(\theta_i,\phi_j)$ accounts for the remaining information and is assumed to be independent and follow $\mathcal{N}(0,v^2(\theta_i,\phi_j))$. The harmonic coefficients $(z_t^{(r)})_{\ell,m}$ are computed by the SHT given by~\cite{Chowdhury:2024}:
\begin{equation}
(z_t^{(r)})_{\ell,m} = \int_{\theta=0}^\pi \int_{\phi=0}^{2\pi} Z_t^{(r)}(\theta,\phi) \overline{Y_{\ell,m}(\theta,\phi)} \, \sin\theta \mathrm{d}\theta \mathrm{d}\phi,
    \label{eq:SHT}
\end{equation}
which can be re-formulated by dropping the dependence on $t$ and $r$ for succinct representation as
\begin{equation}
    (z)_{\ell,m} = \int_{\theta=0}^\pi G_m(\theta) Y_{\ell,m}( \theta, 0 )  \sin\theta  \mathrm{d} \theta \label{coeff2},
\end{equation}
with $Y_{\ell,m}( \theta, \phi ) = Y_{\ell,m}( \theta, 0 ) e^{\mathrm{i}m\phi}$ and 
\begin{equation*}
G_m(\theta)  = \int_{\phi=0}^{2\pi} Z( \theta, \phi ) e^{-\mathrm{i}m\phi} \mathrm{d} \phi = 2\pi \sum_{\ell=|m|}^{L-1} (z)_{\ell,m} Y_{\ell,m}(\theta, 0),
\end{equation*}
which can be used to represent the data in the form
\begin{equation}
Z(\theta,\phi)  = \sum_{m=-(L-1)}^{L-1} G_m(\theta) e^{-\mathrm{i}m\phi}
\label{Eq:isht2}
\end{equation}
with an assumption that the data are band-limited at degree $L$. Noting the relationship between Wigner-$d$ function, denoted by $d^{\ell}_{m,n}(\theta)$ for degree $\ell$ and orders $|m|,|n|\le\ell$, and the spherical harmonic~\cite{Kennedy-book:2013}: 
    $Y_{\ell,m}( \theta, 0 ) = \sqrt{\frac{2 \ell + 1}{4 \pi}} d^{\ell}_{m,0}(\theta)$,
and the expansion of Wigner-$d$ function in terms of complex exponentials~\cite{Kennedy-book:2013}:
$    d^{\ell}_{m,0}(\theta) = \mathrm{i}^{-m} \sum_{m'=-\ell}^\ell d^{\ell}_{m',0}(\frac{\pi}{2}) \,\,d^{\ell}_{m',m}(\frac{\pi}{2}) \,\,
    e^{\mathrm{i}m'\theta},$
we express $G_m(\theta)$ after changing the order of summation as 
\begin{equation}
G_m(\theta)  =  \sum_{m'=-(L-1)}^{L-1} K_{m,m'} \,\,
    e^{\mathrm{i}m'\theta},
    \label{Eq:Gm3}
\end{equation}
where $K_{m,m'}    = \mathrm{i}^{-m} \sum_{\ell=\max(|m'|,|m|)}^{L-1}  \sqrt{(\pi)(2 \ell + 1)} \,\, f_{\ell,m} \times \\
 \quad \quad\quad   d^{\ell}_{m',0}(\frac{\pi}{2}) \,\,d^{\ell}_{m',m}(\frac{\pi}{2})$. Substituting \eqref{Eq:Gm3} in \eqref{coeff2} yields 
\begin{align}
    (z)_{\ell,m} &=    
    \sum_{m''=-(L-1)}^{L-1} \hspace{-3mm} S_{\ell,m,m''}  \hspace{-3mm} \sum_{m'=-(L-1)}^{L-1}  \hspace{-3mm} K_{m,m'} \,I(m'+m'')
    \label{coeff3}
\end{align}
where
$S_{\ell,m,m''} = \mathrm{i}^{-m}  \sqrt{\frac{2 \ell + 1}{4 \pi}}  d^{\ell}_{m'',0}\left(\frac{\pi}{2}\right)  \,\,d^{\ell}_{m'',m}\left(\frac{\pi}{2}\right)$,
and $I(q)$ is an integral of the form
\begin{equation}
I(q) = \int_{\theta \in [0, \pi]} e^{\mathrm{i}q\theta}  \sin\theta  \mathrm{d} \theta = \begin{cases} \delta_{|q|,1}\,\frac{\mathrm{i} q \pi}{2},\, & q\,\,{\rm odd,} \\ \frac{2}{1-q^2}, & q\,\,{\rm even,}
\end{cases}
\end{equation}
where $\delta_{|q|,1}$ is a Kronecker delta function. 

Since the representation of the data in \eqref{Eq:isht2} indicates that $Z(\cdot,\phi)$, with complex exponentials $\{e^{\mathrm{i}m\phi}\}$ as basis functions, is band-limited at $L$, $G_m(\theta)$ can be recovered accurately using the Fast Fourier Transform (FFT) if $N_\phi\ge 2L-1$. We extend the domain of $G_m(\theta)$ to include the points along co-latitude in $(\pi,2\pi)$ for $\theta \in (0,\pi)$ as
$G_m(2\pi-\theta) = 2\pi \sum_{\ell=|m|}^{L-1} f_{\ell,m} Y_{\ell,m}(\theta, \pi) 
    = (-1)^m \,G_m(\theta),$
where we have used $(-1)^m  Y_{\ell,m}(\theta, \pi) =  Y_{\ell,m}(\theta, 0)$. We use \eqref{Eq:Gm3} to recover $K_{m,m'}$ exactly by employing inverse FFT provided $G_m(\theta)$ over $2N_\theta-2$ points along extended co-latitude if $N_\theta >L$. Once we have $K_{m,m'}$ for all $|m|,|m'|<L$, we use \eqref{coeff3} for the exact computation of SHT under the assumption that the data are band-limited. We note that the SHT can be computed with sufficient accuracy using the proposed approach even when the data are not band-limited, as illustrated in \cite{Song:2023}, and the error captured by $\varepsilon_t^{(r)}$ is incorporated in the emulation process.

\subsubsection{Efficient Implementation}
For efficient implementation of SHT, we pre-compute the Wigner-$d$ matrix, thus eliminating the need to recalculate them with each temporal observation. The computation of $G_m(\theta)$ and $K_{m,m'}$ is optimized to $\mathcal{O}(L^2 \log L)$ through the FFT, whether along $\phi$ or an extended $\theta$. Once $K_{m,m'}$ is determined for all $|m|, |m'|<L$ for each observation, the computational complexity of the inner summation in \eqref{coeff3} is $\mathcal{O}(L^3)$ for all $m$ and $m''$. The subsequent outer summation also requires $\mathcal{O}(L^3)$ per observation, resulting in a total complexity of $\mathcal{O}(TL^3)$ for all temporal observations. The Wigner-$d$ functions, specifically for the fixed argument $\pi/2$, can be efficiently calculated using recursion relations, with an $\mathcal{O}(\ell^2)$ complexity for computing $d^{\ell}_{m'',m}(\frac{\pi}{2})$ from $d^{\ell-1}_{m'',m}(\frac{\pi}{2})$, resulting in a total of $\mathcal{O}(L^3)$ across all degrees and orders. The calculation of $S_{\ell,m,m'}$ also takes $\mathcal{O}(L^3)$. As $S_{\ell,m,m'}$ does not depend on the data, it can be pre-computed for $\ell<L$, $|m|\le \ell$, and $|m'|<L$, necessitating $\mathcal{O}(L^3)$ in storage space. While our pre-computation strategy does not change the asymptotic computational complexity of SHT, it significantly reduces actual computation time. Additionally, we underscore that our proposed SHT computation technique is fully parallelizable, offering a linear computational complexity of $\mathcal{O}(L)$ for computing SHT for different time points simultaneously.

\subsubsection{Modeling the Temporal Dependence and Constructing a Covariance Matrix}
After the computation of SHT of the stochastic component for all time points, we can form a matrix $\mathbf{F}^{(r)}=(\mathbf{f}^{(r)}_1,\ldots,\mathbf{f}^{(r)}_T)\in\mathbb{R}^{L^2\times T}$ where  $\mathbf{f}^{(r)}_t=((z_t^{(r)})_{00},\ldots,(z_t^{(r)})_{LL})^\top\in\mathbb{R}^{L^2}$ is a vector of spherical harmonic coefficients for each time slot $t=1,\ldots,T$. We model the temporal dependence using a vector auto-regressive model of order $P$, that is,
 $   \mathbf{f}^{(r)}_t= \sum_{p=1}^P\Phi_p\mathbf{f}^{(r)}_{t-p}+\bm{\xi}^{(r)}_t,$
where each $\Phi_p\in\mathbb{R}^{L^2\times L^2}$, $p=1,2,\hdots,P$, is assumed to be a diagonal correlation matrix~\cite{Song:2023}, and $\bm\xi_t^{(r)}\overset{i.i.d.}{\sim}\mathcal{N}_{L^2}(\mathbf{0},\mathbf{U})$. The covariance matrix $\mathbf{U}$ can be empirically evaluated with computational complexity $\mathcal{O}(L^4T)$ as  
\begin{equation}
    \hat{\mathbf{U}} = \frac{1}{R(T-P)}\,\sum_{r=1}^R\,\sum_{t=P+1}^T{\bm{\xi}^{(r)}_t}\,\big({\bm{\xi}^{(r)}_t}\big)^\top,
\end{equation}
and then factorized via Cholesky decomposition with computational complexity $\mathcal{O}(L^6)$ as $\hat{\mathbf{U}}=\mathbf{V}\mathbf{V}^\top$. In cases where $R(T-P) < L^2$, we introduce a minor perturbation along the diagonal of $\hat{\mathbf{U}}$ to ensure it remains positive definite. This completes the design of our emulator capable of generating emulations using the matrices $\mathbf{V}$, $\{\Phi_p\}_{p=1}^P$ and parameters~(or matrices) $v$, $\sigma,\beta_0,\beta_1,\beta_2,\rho,\{a_k,b_k\}_{k=1}^{K}$ defining the characteristics of spatio-temporal climate data.  

\subsection{Generating Emulations}

Using $\mathbf{V}$, we first generate $\bm\xi_t^{(r)}\overset{i.i.d.}{\sim}\mathcal{N}_{L^2}(\mathbf{0},\mathbf{U})$ and determine $\mathbf{f}^{(r)}_t$ by employing $\Phi_p$ for $p=1,\hdots,P$ in $\mathcal{O}(L^2T)$. We then compute inverse SHT of the spherical harmonic coefficients $\mathbf{f}^{(r)}_t$ to obtain $(\tilde Z_t^{(r)})$ over a grid of $N_\theta$ latitudes and $N_\phi$ longitudes with the computational complexity of $\mathcal{O}(L^3T)$. Add $\varepsilon_{t}^{(r)}\sim\mathcal{N}(0,v^2)$ to $(\tilde Z_t^{(r)})$ to get $(Z_t^{(r)})$. Finally, we use \eqref{eq:decouple} to generate emulations with the overall computational complexity of $\mathcal{O}(L^3T)$.

\subsection{Using \parsec to Meet Exascale Challenges}

Scale changes everything, escalating minor issues to major performance bottlenecks. Scaling up our experiments, especially on platforms where each node has a tremendous computing capability, highlighted two potential issues: the importance of ordering communications, especially collective communications, and the significance of minimizing memory waste. \parsec has support for both of these capabilities, including user-driven support via annotations in the source code, but this support was not previously tested at this scale.

Tiles from a regularly distributed matrix with varied precision require different amounts of storage. \parsec has support for dynamic memory allocations based on the incoming data ~\cite{ipdps2021}, but as the communication layer can reshape the data to adapt it to the precision required by the successors' tasks, this allocation needs to be set dynamically, driven by the sender. Moreover, this dynamic allocation should be able to handle data on both CPU and accelerator memory.

Collective communications in \parsec are conceptually similar to those in MPI, with the major difference that no group or communicator is known in advance; the group will be dynamically built based on the location of successor tasks and propagated (either physically or symbolically) with the collective message. This means that each node of the propagation tree will locally construct the group of collectives and identify the processes with which they interact before participating in the collective communication. Moreover, as many collective communications are ongoing simultaneously, the original version of \parsec tried to maximize the overall bandwidth and allow for multiple concurrent collective communications. In other words, \parsec favored bandwidth over latency, resulting in individually longer collective latency but an overall better bandwidth usage. At scale, especially for strong scaling experiments, such a strategy is sub-optimal, as it creates starvation points. Thus, we realigned the priorities of the collective communications to focus on individual latency first, leading to faster and more regular time-to-solution.

\subsection{Mixed-Precision Computation Innovations}
Recent GPUs from various vendors are designed to deliver high performance at lower precision levels, catering to modern AI workloads. In earlier research, we introduced mixed-precision (MP) computation for large-scale single matrices, utilizing tile-based algorithms to distribute precision across different tiles according to application features. For example, precision distribution can be based on bandwidth, as seen in band-based MP computations \cite{abdulah2019geostatistical, abdulah2021accelerating}, or it can be tailored around numeric features, as demonstrated in tile-centric MP approaches \cite{cao2022reshaping}. All these studies were around CPU-based systems, including Intel, ARM, and Fujitsu A64FX Systems. 

Recently, we expanded this approach to single and multiple GPUs \cite{cao2023reducing}, achieving improved performance and reduced power consumption for environmental applications. This advancement aims to enhance the accuracy of predictions using existing geospatial data. Here, we broaden the capabilities of the \parsec runtime system and mixed-precision computation using tile-based linear algebra algorithms to large-scale GPU-based systems, specifically, Frontier (AMD MI250X), Alps (NVIDIA H100), Leonardo (NVIDIA A100), and Summit (NVIDIA V100), and  we supplement our study with efficiency and scalability assessment results for climate emulators.



%% file: measures.tex
\ifthenelse {\boolean{GuideOn}} {
\textbf{(Note that preference is given to performance actually measured [not projected],
based on the entire application [including I/O] and with uniform precision.
Explain in detail if any portion of the total runtime was not included in the
measurements, if and where different precisions were used, or any attributes
listed in Section 3 as “other”).\\
what application(s) was used to measure performance (1 p max)\\
system and environment where performance was measured (1 p max)\\
}
} {
}



\begin{figure*}[!ht]
\vspace{-2.5mm}
    \begin{subfigure}[b]{0.23\linewidth}
        \centering        \includegraphics[width=\linewidth,trim = 0.8cm 2.5cm 0.8cm 2.5cm]{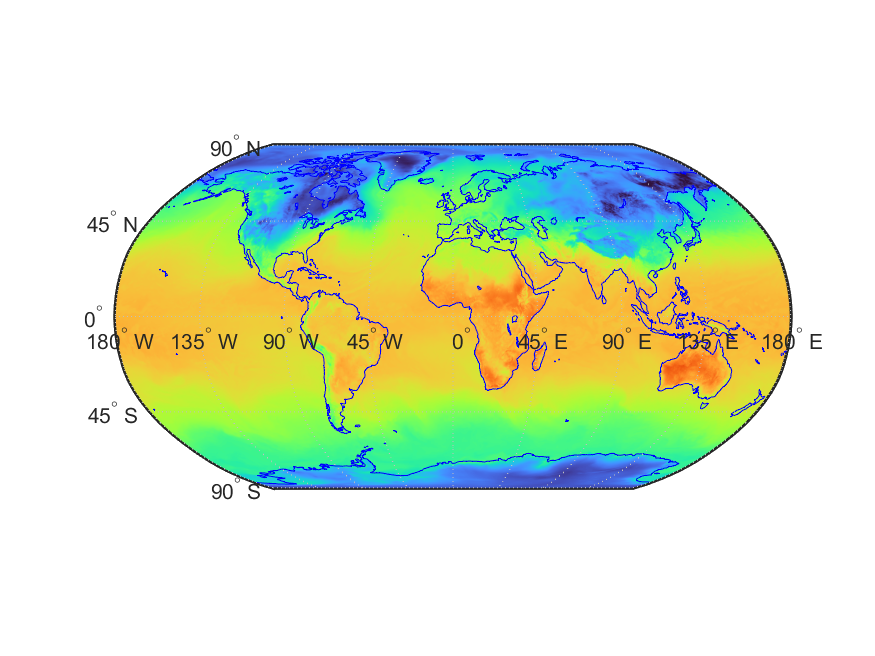}
       \caption{{\small ERA5 Data, Jan. 01, 2019}}
        \label{fig:subfig1}
    \end{subfigure}
    \begin{subfigure}[b]{0.23\linewidth}
        \centering
        \includegraphics[width=\linewidth,trim = 0.8cm 2.5cm 0.8cm 2.5cm]{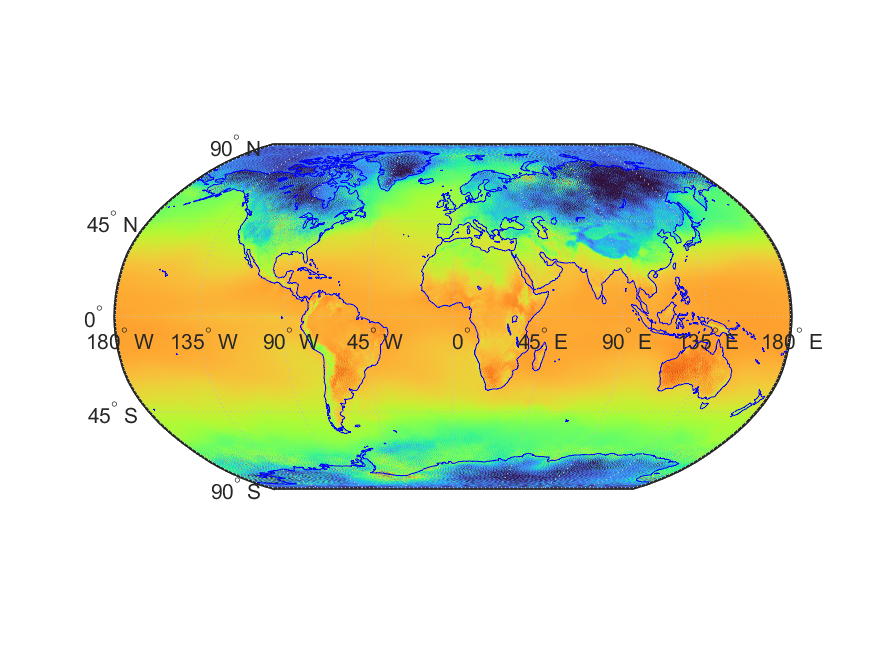}
       \caption{{\small Emulation (DP)}}
        \label{fig:subfig2}
    \end{subfigure}
    \begin{subfigure}[b]{0.23\linewidth}
        \centering
        \includegraphics[width=\linewidth,trim = 0.8cm 2.5cm 0.8cm 2.5cm]{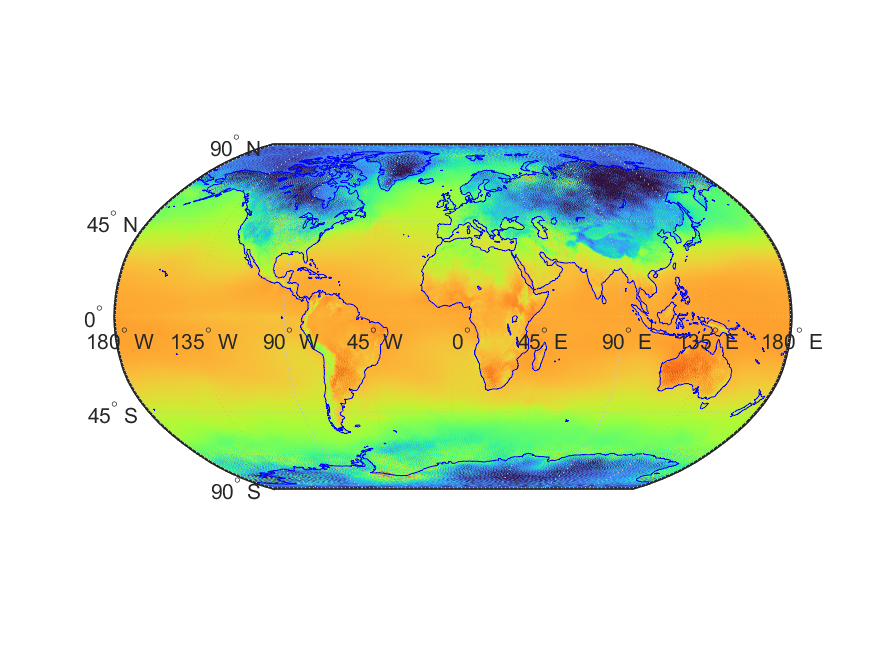}
       \caption{{\small Emulation (DP/SP)}}
        \label{fig:subfig3}
    \end{subfigure}
    \begin{subfigure}[b]{0.23\linewidth}
        \centering
        \includegraphics[width=\linewidth,trim = 0.8cm 2.5cm 0.8cm 2.5cm]{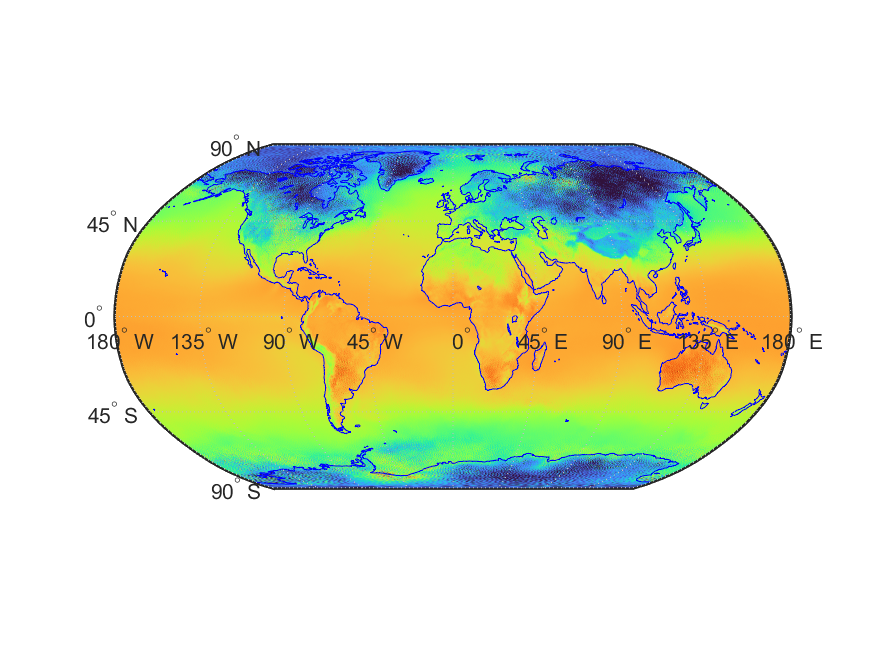}
       \caption{{\small Emulation (DP/HP)}}
        \label{fig:subfig4}
    \end{subfigure}
    \\[7.5mm]
    \begin{subfigure}[b]{0.23\linewidth}
        \centering
        \includegraphics[width=\linewidth,trim = 0.8cm 2.5cm 0.8cm 2.5cm]{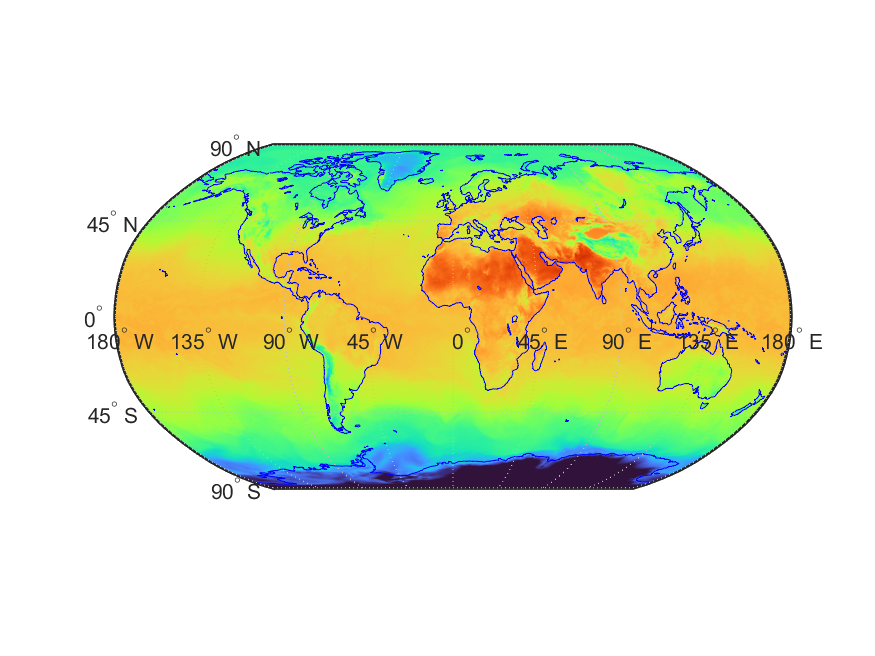}
       \caption{{\small ERA5 Data, Jun. 01, 2019}}
        \label{fig:subfig5}
    \end{subfigure}
    \begin{subfigure}[b]{0.23\linewidth}
        \centering
        \includegraphics[width=\linewidth,trim = 0.8cm 2.5cm 0.8cm 2.5cm]{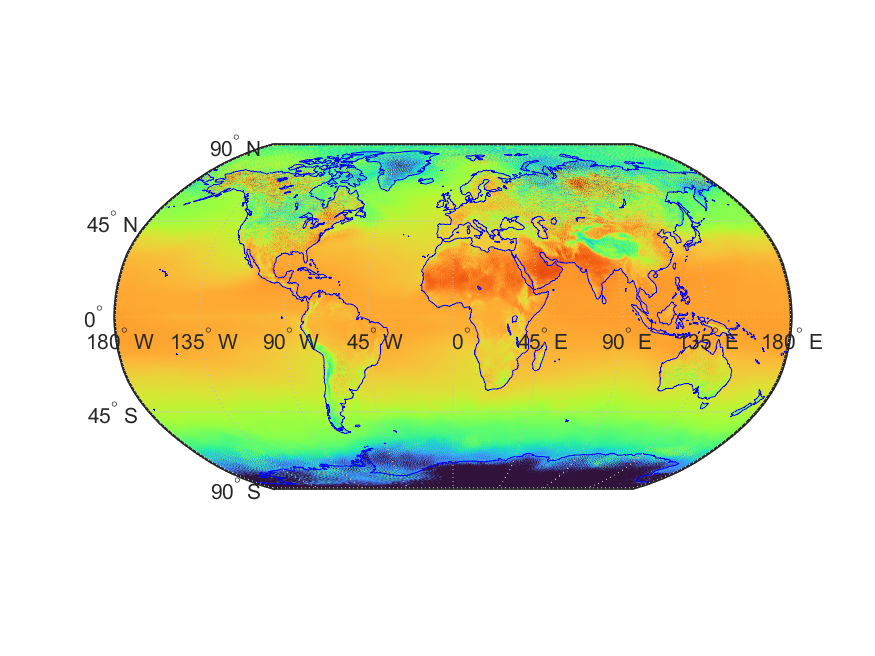}
       \caption{{\small Emulation (DP)}}
        \label{fig:subfig6}
    \end{subfigure}
    \begin{subfigure}[b]{0.23\linewidth}
        \centering
        \includegraphics[width=\linewidth,trim = 0.8cm 2.5cm 0.8cm 2.5cm]{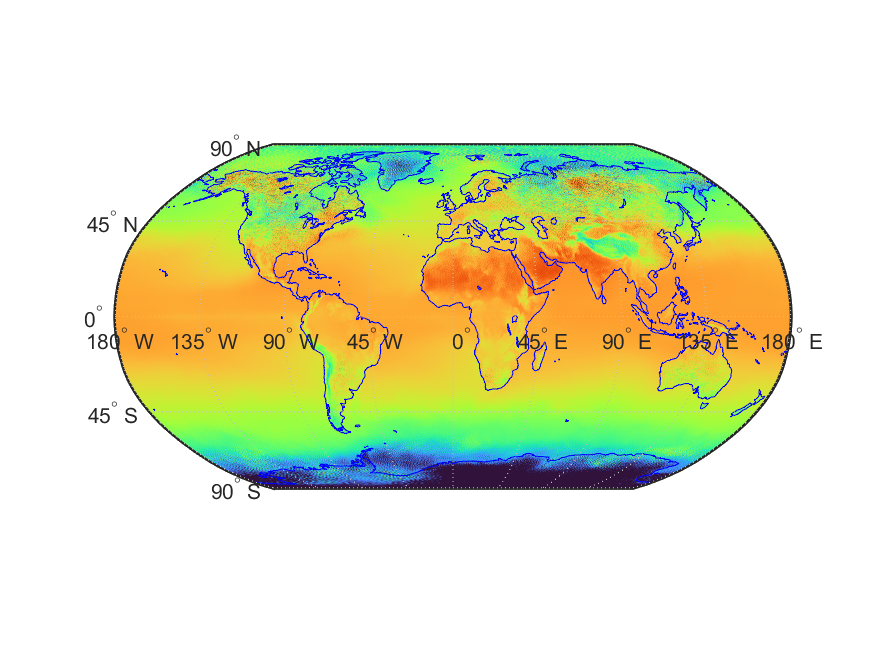}
       \caption{{\small Emulation (DP/SP)}}
        \label{fig:subfig7}
    \end{subfigure}
    \begin{subfigure}[b]{0.23\linewidth}
        \centering
        \includegraphics[width=\linewidth,trim = 0.8cm 2.5cm 0.8cm 2.5cm]{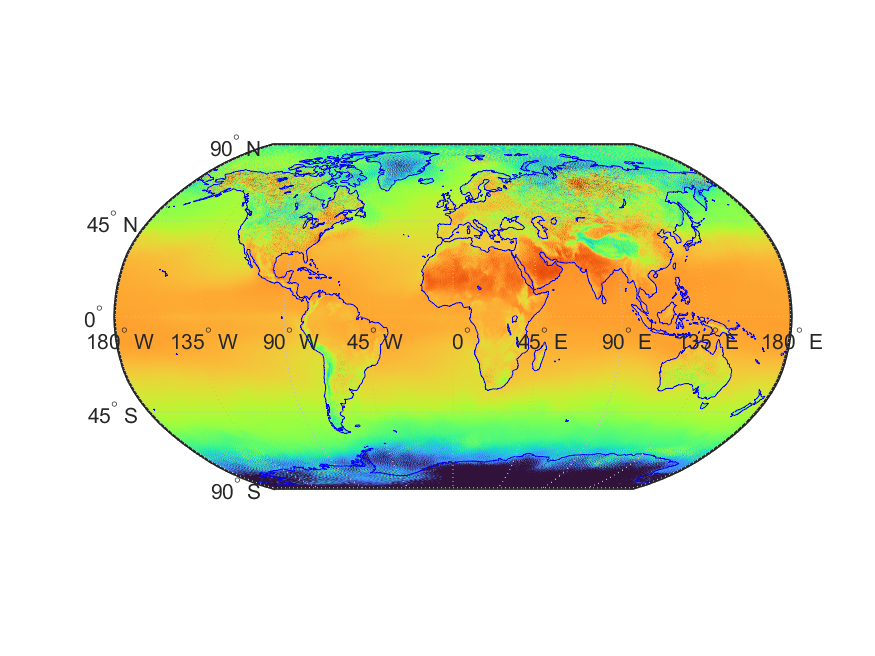}
       \caption{{\small Emulation (DP/HP)}}
        \label{fig:subfig8}
    \end{subfigure}
    \hspace{-0.01cm}
    \begin{minipage}[b]{0.04\textwidth}
        \vspace{-11cm}
        {\includegraphics[scale=0.28,trim = 1.2cm 6.5cm 0.5cm 0cm]{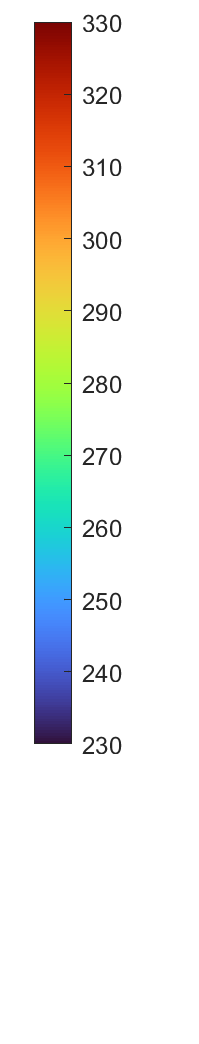}}
      \end{minipage}
      \vspace{-1mm}
    \caption{{\small ERA5 surface temperatures (in K) on (a) Jan. 1, 2019, and (e) Jun. 1, 2019. The corresponding emulated temperatures using DP, DP/SP, and DP/HP, are plotted in the same row.}}
    \vspace{-5mm}
    \label{fig:emulation_maps}
\end{figure*}


\subsection{Spatio-temporal Dataset, Upsampling and Accuracy}
We consider the surface temperature measured at 2 meters above the surface, encompassing land, sea, and inland water bodies, as spatio-temporal climate data. We use the ERA5 dataset, which provides the surface temperature spanning the last 83 years (1940-2022) at a spatial resolution of $0.25^{\circ}$ corresponding to $N_\theta=721$ latitude points and $N_\phi= 1440$ longitude points and the spherical harmonic band-limit $L=720$. In anticipation of future climate simulations and experiments expected to feature even higher spatial resolutions, we perform spline interpolation to upscale the data to higher spatial resolutions consistent with the band-limits of \bandlimits (see green stars in Figure~\ref{fig_overview}). This upscaling of the number of grid points serves dual purposes: i) to demonstrate the scalability of the design and computational performance of our emulator and ii) to ensure our approach remains relevant and aligned with advancements in climate modeling. We use $K=5$ to capture periodic variations along the time and $P=3$ for auto-regressive modeling of spherical harmonic coefficients, a choice deemed sufficient based on existing related research \cite{Huang:2023,Song:2023}. For the temporal dimension, we consider hourly and daily resolutions, resulting in $T = 306{,}600$ (35 years, hourly, 1988-2022) and $T = 30{,}295$ (83 years, daily 1940-2022) time slots, after adjusting for the omission of an extra day in leap years. As highlighted earlier, even daily resolution has not been fully adopted in the existing emulator designs primarily due to the significant computational demand to process large amounts of data. We plot the hourly ERA5 data alongside the emulations in Figure~\ref{fig_illustraion} for two days. We also plot the daily simulations and emulations for the same two days in Figure~\ref{fig:emulation_maps} for different mixed-precision variants. These figures illustrate the statistical consistency~(studied in detail in \cite{Song:2023}) between the emulations and simulations at different precision levels employed in the tile-based decomposition of the covariance matrix.

\subsection{Performance Measured in Flops/s}
We assess the numerical robustness of our mixed-precision approach by applying the climate emulator with various $L$ values, specifically, \bandlimits. We report the Flops/s and scalability achieved on up to \frontierNode nodes on Frontier (\frontierGPU AMD MI250X GPUs), \ALPSNode nodes on Alps (\ALPSGPU A100 GPUs), \leonardoNode nodes on Leonardo (\leonardoGPU A100 GPUs), and \summitNode nodes of Summit (\summitGPU V100 GPUs). To optimize performance, we fine-tuned our code and integrated it with the highly optimized BLAS/LAPACK libraries available on each of these systems.

We conducted experiments on the four supercomputers, utilizing four precision variants: full double-precision (DP); a single band as DP with the remaining tiles as SP (DP/SP); DP with 5\% as SP and the remaining tiles as HP (DP/SP/HP); and a single band as DP with the remaining tiles as HP (DP/HP). The performance results were reported in terms of achieved Flops/s for a single Cholesky factorization, which is required after the SHT step to emulate new climate datasets. We varied the problem sizes to encompass all spatial resolutions down to \resolutionDegree. The performance data spanned different numbers of nodes across the various systems, limited by availability but nevertheless providing a comparison of our computational capability across different configurations and resolutions. This approach allowed us to assess the efficiency of each variant under different computational loads and system scales.

\subsection{Strong and Weak Scaling Efficiency}

Having greatest availability on about-to-be-retired Summit, we gave it a triumphant swan song of strong and weak scaling experiments using up to $2{,}048$ nodes ($12{,}288$ NVIDIA V100 GPUs) with four Cholesky decomposition variants: the reference DP, and variants DP/SP, DP/SP/HP, and DP/HP. 
For the weak scaling experiments, we assigned the same amount of data to each GPU while varying the number of GPUs. This approach helped us evaluate how effectively our MP code can maintain performance per GPU as the scale of the system increases. It allowed us to assess the consistency of performance across different GPU counts, to understand how the application scales horizontally.

For strong scalability, we used a constant workload that $512$ nodes of Summit would typically handle and tested this workload on larger configurations of $1{,}024$ and $2{,}048$ nodes. This allowed us to measure the scaling efficiency across the different precision variants. Specifically, we examined how the performance per GPU with the fixed workload decreased with adding more GPUs and computed the efficiency as the performance ratio per GPU on the 512-node configuration. This metric helps in understanding the effectiveness of resource utilization and in pinpointing potential overheads or inefficiencies in the parallelization of the code, especially as we scale up to a larger number of GPUs.

\subsection{Attributes of the HPC Systems}
This study focuses on four GPU-based systems. 
Frontier at ORNL consists of $9{,}472$ nodes, each equipped with four AMD MI250X GPUs; each GPU is a multi-chip module (MCM) with two AMD Graphics Compute Dies (GCDs) and four Terabytes of flash memory. Frontier has a theoretical peak performance of $1.71$ EFlop/s in DP operations. Recently listed Alps at CSCS consists of many resources; we focus on the Grace-Hopper partition of $2{,}688$ CPU-GPU supernodes, each equipped with four NVIDIA GH200 with 72 ARM cores CPU and one NVIDIA H100 Tensor Core GPU with 96GB of HBM3. Alps has a theoretical peak performance of $353.75$ PFlop/s in DP operations.
Leonardo at CINECA comprises $3{,}456$ nodes, each with four NVIDIA A100 64GB GPUs and boasting a theoretical peak performance of $306.31$ PFlop/s in DP. Summit at ORNL supercomputer consists of $4{,}608$ compute nodes, each equipped with two 22-core IBM Power9 CPUs and six NVIDIA Tesla V100 GPUs, with over 600 GB of coherent memory accessible by both CPUs and GPUs and a theoretical peak performance of $200.79$ PFlop/s in DP.


%% file: performance.tex

\ifthenelse {\boolean{GuideOn}} {
\textbf{include scalability (weak and strong), time to solution, efficiency (of bottleneck resources),
and peak performance (2 pp max)\\}
} {
}

\begin{figure*}[!t]
\vspace{-2mm}
\centering
    \begin{minipage}[b]{0.49\textwidth}
        \centering
        \includegraphics[width=0.75\linewidth]{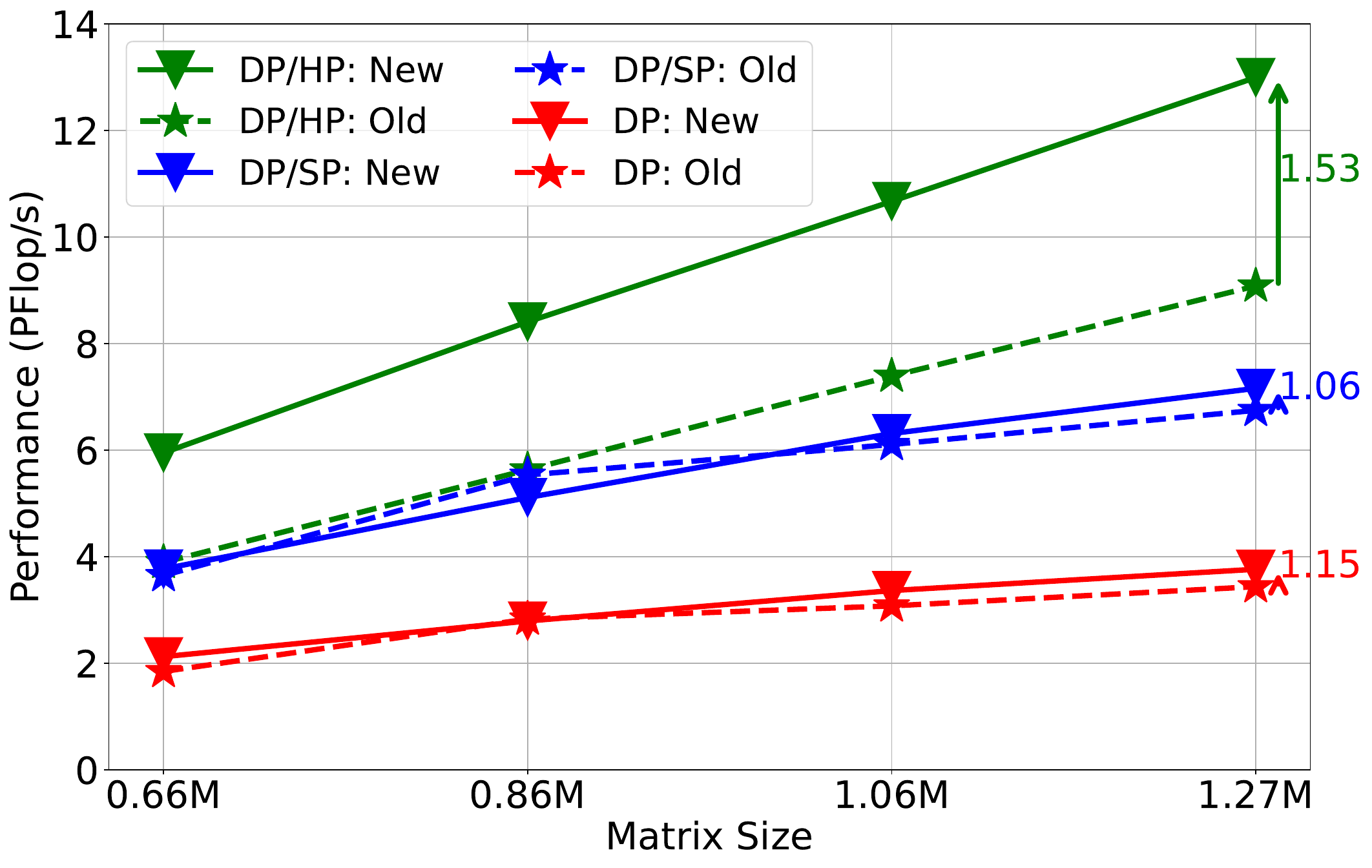}
\caption{{\small
 Performance of Cholesky factorization for DP, DP/SP, and DP/HP  on 128 nodes of Summit ($768$ NVIDIA V100 GPUs) compared to the work in~\cite{abdulah2021accelerating}, with varying covariance matrix sizes.}}
\label{fig:perf-128summit}
    \end{minipage}
    \hfill
    \begin{minipage}[b]{0.49\textwidth}
        \centering
        \includegraphics[width=0.75\linewidth]{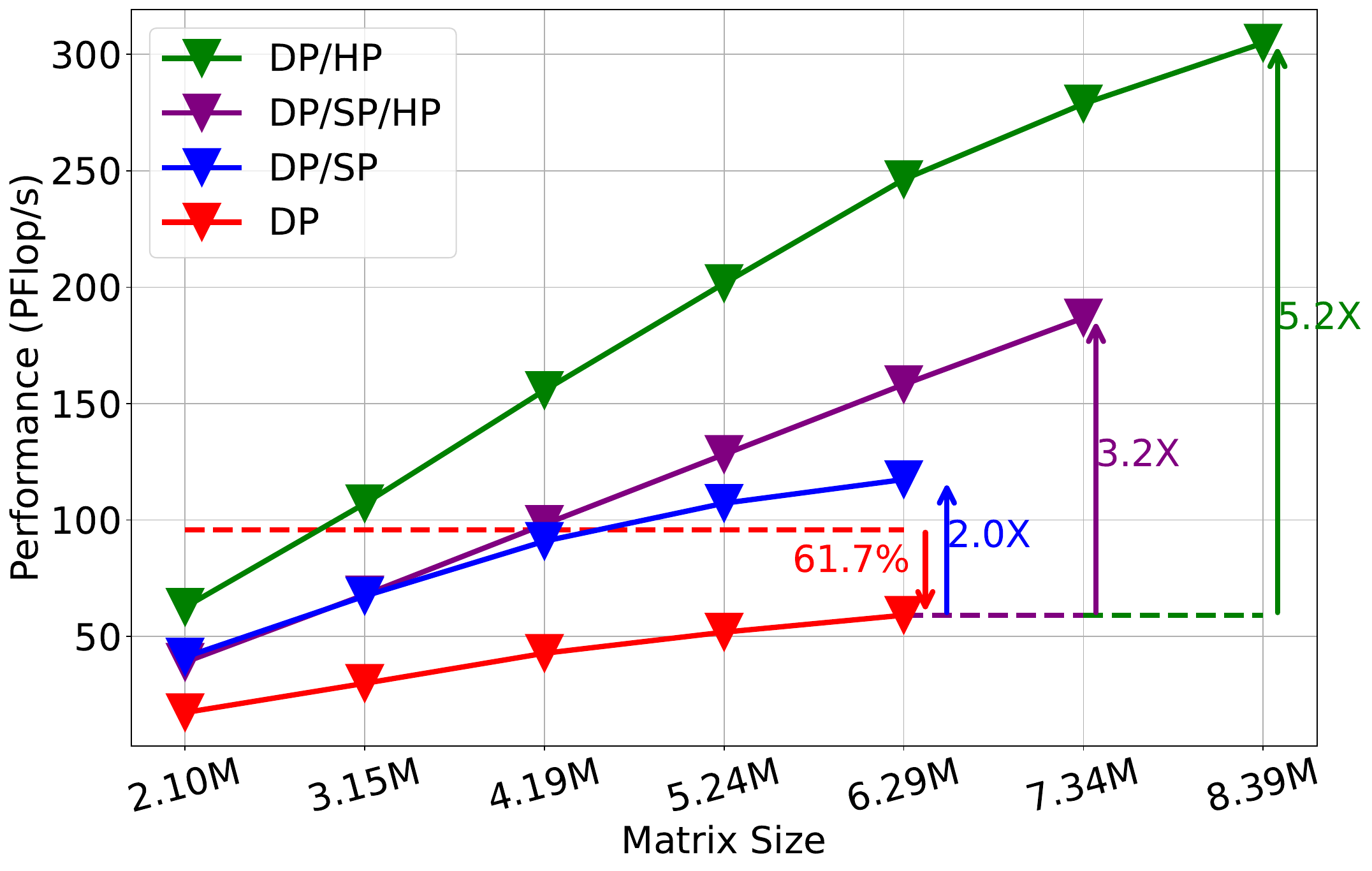}
\caption{{\small
 Performance of Cholesky factorization using four precision levels: DP, DP/SP/HP, DP/SP, and DP/HP  on $2{,}048$ nodes of Summit ($12{,}288$ NVIDIA V100 GPUs), with varying covariance matrix sizes.}}        \label{fig:perf-summit_2048}
    \end{minipage}
\end{figure*}

\begin{figure*}[h]
    \vspace{-2mm}
    \centering
    \begin{subfigure}{0.48\textwidth}
        \centering        \includegraphics[width=0.75\textwidth]{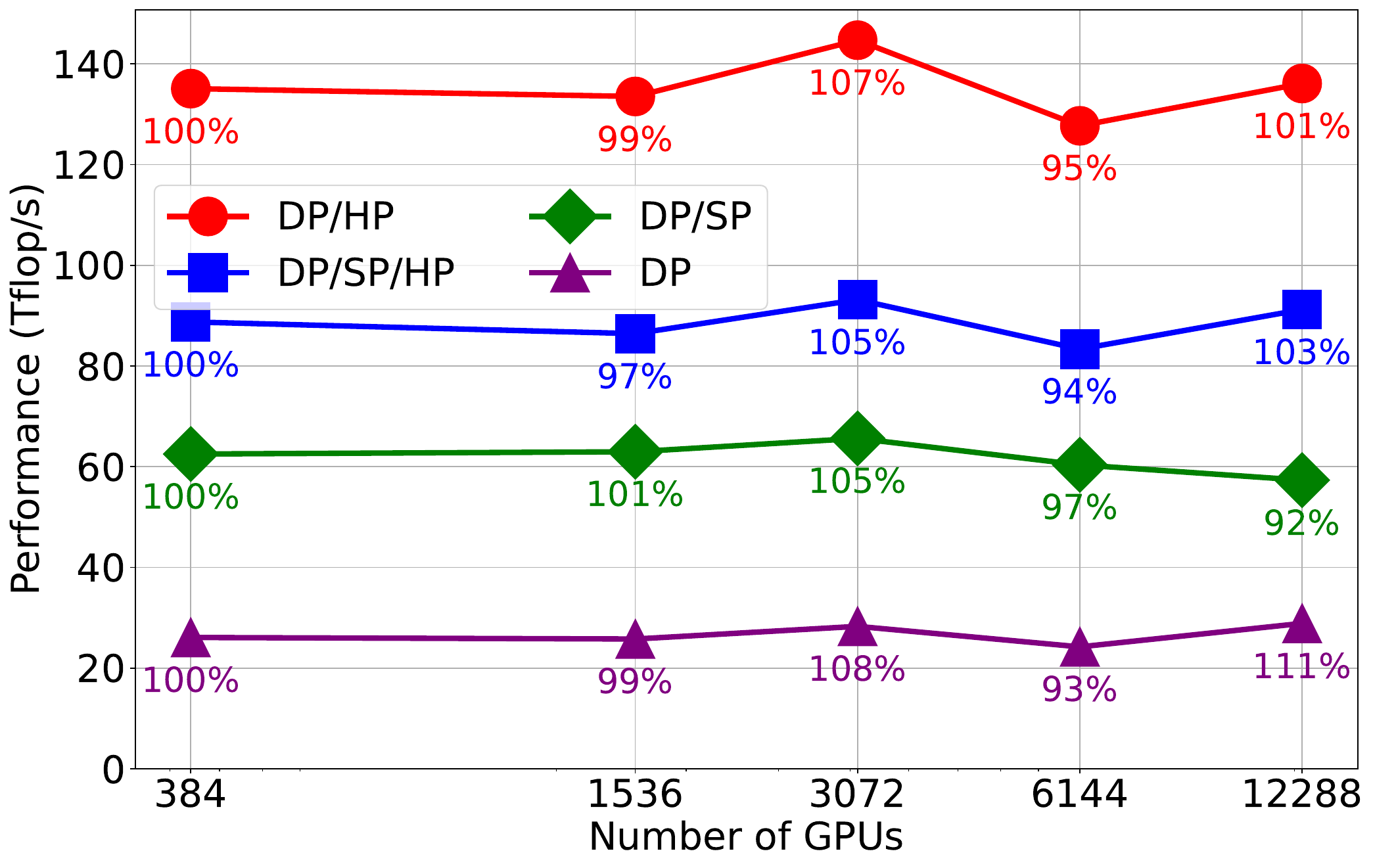}
    \end{subfigure}
    \hspace{4mm}
    \begin{subfigure}{0.48\textwidth}
        \centering        \includegraphics[width=0.75\textwidth]{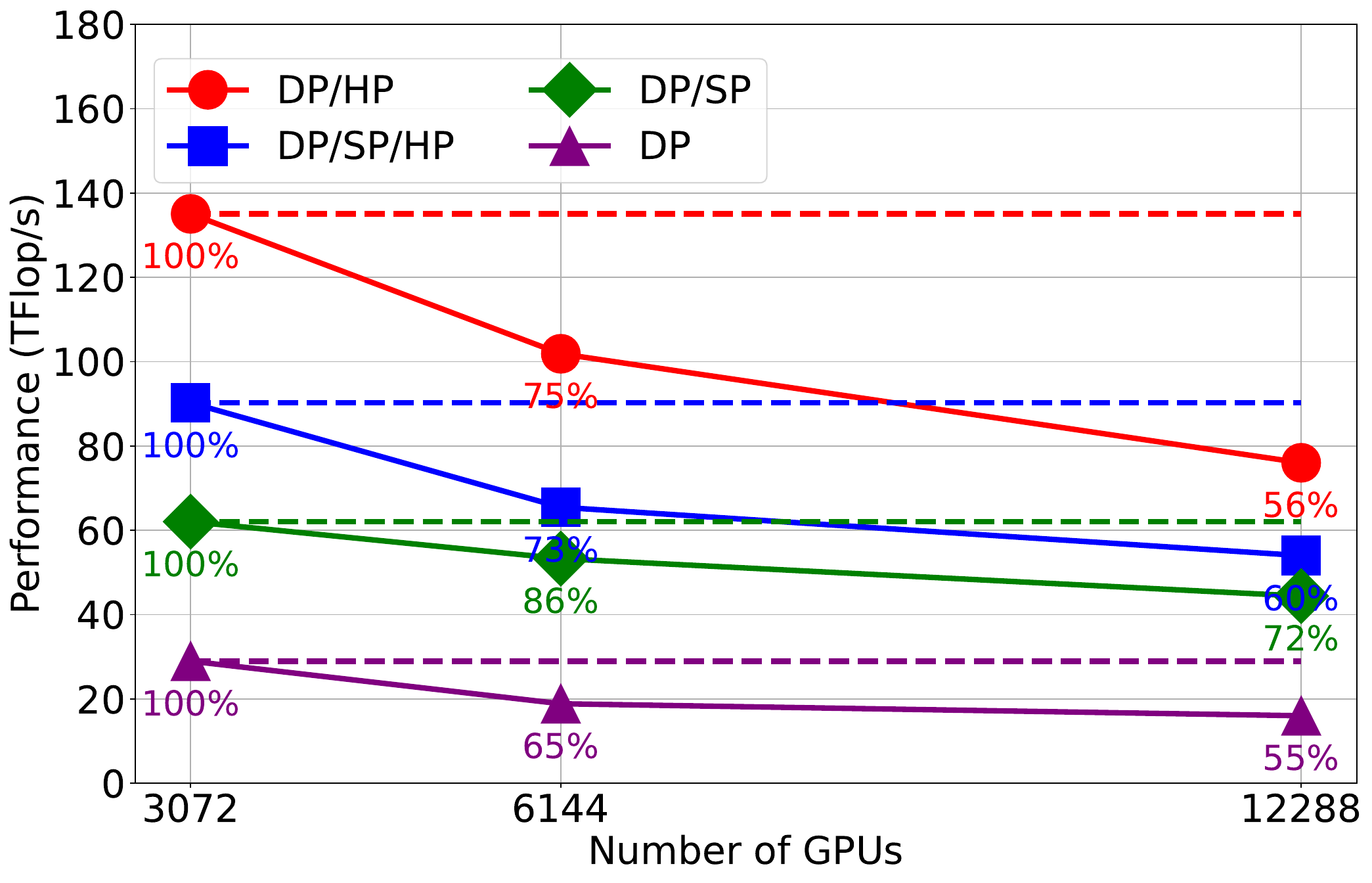}
    \end{subfigure}
    \vspace{-1mm}
    \caption{{\small Weak (left) and strong (right) scaling on $2{,}048$ nodes of Summit ($12{,}288$ NVIDIA V100 GPUs), using various precision levels, i.e., DP, DP/SP, DP/SP/HP, and DP/HP.}}
    \vspace{-4mm}
    \label{fig:perf-summit_strong_weak}
\end{figure*}

We analyze performance on the four above-mentioned systems, covering various problem sizes and weak and strong scalability efficiency on Summit.

\subsection{Performance of Mixed-Precision Cholesky on Summit}
In~\cite{abdulah2021accelerating}, we propose a banded-mixed-precision approach through PaRSEC at large-scale. At this scale, tile-based mixed-precision Cholesky factorization has a special communication pattern. First, POTRF ($c, c$) on the diagonal tile broadcasts to TRSMs in column $c$ for a given tile-based matrix. Secondly, TRSM ($r, c$) broadcasts to GEMMs in row $r$, GEMMs in column $c$, and SYRK ($r, c$). Due to different precisions, a task might receive a tile at a different precision than it operates on, necessitating a precision conversion. This conversion can occur at the sender or the receiver. If we apply down-precision conversion at the sender, we boost performance by reducing communication. Send-based conversion enhances performance, as exemplified by TRSM to GEMMs conversions, which reduces repeated conversions across successive GEMMs. An evaluation with single and multiple GPUs of sender/receiver-based conversions was introduced in~\cite{cao2023reducing}. In this work, we have optimized the performance by dynamically applying sender-based conversion and compared the performance to our previous work~\cite{abdulah2021accelerating} on 128 nodes of Summit. As shown in Figure~\ref{fig:perf-128summit}, we can obtain a speedup up to 1.53X with DP/HP. The other two curves have slightly improved, i.e., $1.15$X and $1.06$X for DP and DP/SP, respectively,  because the advantage in communication is hindered by the computation overhead, resulting in a negligible performance gain compared to the DP/HP variant.



Figure~\ref{fig:perf-summit_2048} shows the performance comparisons of several precision configurations: DP, DP/SP, DP/SP/HP, and DP/HP on $2{,}048$ nodes of Summit. The red dashed line represents the peak performance of DP on $2{,}048$ Summit nodes. The figure illustrates that DP Cholesky achieves 61.7\% of the peak performance. The figure also reports the speedups achieved by the three MP configurations compared to DP computation, showing speedups of $2$X, $3.2$X, and $5.2$X for DP, DP/SP/HP, and DP/HP, respectively. The figure also demonstrates that with a problem size of approximately 8M, DP/HP Cholesky on $2{,}048$ nodes can achieve up to $304.84$ PFlops/s.


\subsection{Scalability of Mixed-Precision Cholesky on Summit}

Figure~\ref{fig:perf-summit_strong_weak} (left) illustrates the weak scalability of Summit performing MP Cholesky decomposition. The reported performance is normalized to reflect single-node performance and facilitate a more granular analysis. As the number of nodes increases from $64$ to $2{,}048$, each maintains performance levels per node for different configurations. Figure~\ref{fig:perf-summit_strong_weak} demonstrates excellent scaling efficiency when maintaining the same workload per GPU as the number of GPUs increases. Compared to the baseline performance on $384$ GPUs, the scaling efficiency ranges from 92\% to 111\% across various GPU counts, up to $12{,}288$ V100 GPUs, using different MP variants.


Figure~\ref{fig:perf-summit_strong_weak} (right) illustrates the strong scalability on the Summit system when performing MP Cholesky decomposition. This figure shows how the system handles a fixed total workload as the number of nodes increases from $3{,}072$ to $12{,}288$ GPUs, effectively showcasing the utilization of additional computational resources while maintaining the same problem size. The largest problem size that fits the memory of $3{,}072$ GPUs is used as the fixed workload. The performance metrics are also normalized to show performance per node. As typically expected in strong scalability scenarios, the performance per node tends to decrease when more resources are applied to the same problem size. For instance, with the DP variant, the efficiency drops to about 55\% per node when comparing the performance of $2{,}048$ nodes with $512$ nodes.

However, efficiency improves noticeably when using mixed-precision variants. Efficiency increases to 72\% for DP/SP and 60\% for DP/SP/HP, indicating that these configurations mitigate performance degradation more effectively than the pure DP setup. This suggests that employing mixed-precision can be a strategic choice for enhancing computational efficiency in large-scale systems, especially when significantly scaling up the number of nodes. The efficiency of DP/HP in a large number of nodes has dropped to 56\% due to less work being required per node compared to the DP/SP variant.


\subsection{Performance Comparisons of Mixed-Precision Cholesky} 

Table~\ref{tab:nodes_1k} highlights DP/HP performance in PFlop/s of the mixed-precision Cholesky on $1{,}024$ nodes of the four systems, i.e., Frontier, Alps, Leonardo, and Summit, maxing out the device memory with each matrix size in addition to PaRSEC internal memory buffers. Moreover, since Summit has a higher GPU count per node than other systems, we normalize the absolute performance per GPU. We can observe performance improvements in NVIDIA GPU hardware generations. While NVIDIA A100 achieves similar performance to AMD MI250X, NVIDIA GH200 outperforms it by 1.6X. There is still room for further improvements on Frontier and Alps systems by leveraging their network interconnect using CUDA-aware MPI to mitigate data movement overheads. This requires additional support within PaRSEC and will be addressed in future work.

\subsection{Experiments of Largest Scale Climate Emulators}

Figure~\ref{fig:perf-large-all} reports the largest scale experiments conducted on the Leonardo and Summit systems, using DP/HP configurations with $1{,}024$ and $3{,}072$, respectively. Additionally, we report Alps results using three node counts, i.e., $1{,}024$, $1{,}600$, and $1{,}932$, and Frontier results using four node counts, i.e., $2{,}048$, $4{,}096$, $6{,}400$, $9{,}025$. We achieved $0.243$ EFlop/s on the Leonardo system using $1{,}024$ nodes with a problem size of approximately $8.39$M. On the Summit system, activating $3{,}072$ nodes yielded $0.375$ EFlop/s for a problem size of around $12.58$M. On Alps, we measured $0.364$, $0.623$ and \ALPSPerf EFlop/s on $1{,}024$, $1,600$ and $1{,}936$ nodes with problem sizes of approximately $10.49$M, $14.42$M, and $15.73$M, respectively. Finally, on the Frontier system, we executed on  $2{,}048$, $4{,}096$, $6{,}400$, and $9{,}025$ nodes and achieved $0.316$, $0.523$, $0.715$, and \frontierPerf EFlop/s with problem sizes of approximately $12.58$M, $16.78$M, $20.97$M, and $27.24$M, respectively.

\newcolumntype{Y}{>{\centering\arraybackslash}X}
\begin{table}[!t]
\centering
\caption{DP/HP Performance on $1{,}024$ nodes of the four systems, i.e., Frontier, Alps, Leonardo, and Summit.}
\begin{tabularx}{\linewidth}{|c|Y|Y|Y|Y|}
\hline
\textbf{System} & \textbf{Frontier} & \textbf{Alps} & \textbf{Leonardo} & \textbf{Summit} \\ \hline
Vendor & AMD & \multicolumn{3}{c|}{NVIDIA} \\ \hline
Chip & MI250X & GH200 & A100 & V100 \\ \hline
\# GPUs & $4{,}096$ & $4{,}096$ & $4{,}096$ & $6{,}144$ \\ \hline
Matrix Size & $8.39$M & $10.49$M & $8.39$M & $6.29$M \\ \hline
\makecell{Performance\\(PFflop/s)} & $223.7$ & $384.2$ & $243.1$ & $153.6$ \\
\hline
TFlop/s/GPU & $54.6$ & $93.8$ & $57.2$ & $25$ \\ \hline
\end{tabularx}
\vspace{-4mm}
\label{tab:nodes_1k}
\end{table}

%% file: implications.tex
\ifthenelse {\boolean{GuideOn}} {\textbf{implications for future systems and applications (1 p max)\\
}} {
}
\label{sec:implications}

Dynamic runtime systems have become critical to scalability and performance. Not only do they insulate computational scientists from the programming aspects, allowing them to focus on algorithmics and science, but they also enable computer scientists to quickly explore and adapt solutions to the evolving needs of applications. The level of performance obtained on the highest-end platforms shows that task-based runtime systems have matured, outpacing their initial node-level scope to become scalable partners for HPC applications that can handle resource management and communication overlap, assuming the algorithm has enough computations. 

In this work, we propose a mixed-precision approach that leverages the capability of modern GPUs to perform fast, lower-precision computations with the power of runtime systems. This enables us to develop efficient solvers, such as Cholesky factorization, that can accelerate the performance of applications at their algebraic core. With dynamic runtime systems, we balance load across inhomogeneous tiles of large-scale matrices and hide communication latencies with dense compute-bound algorithms. While this solution is assessed in the context of a climate emulator, it has the potential to benefit a wide range of applications that require rapid computations using dense linear solvers. We have optimized and tested our code across four different GPU architectures -- AMD MI250X, NVIDIA GH200, NVIDIA A100, and NVIDIA V100 -- on four distinct systems:  Frontier, Alps, Leonardo, and Summit.

\begin{figure}[!t]
\vspace{-2mm}
\center
\includegraphics[width=0.775\linewidth]{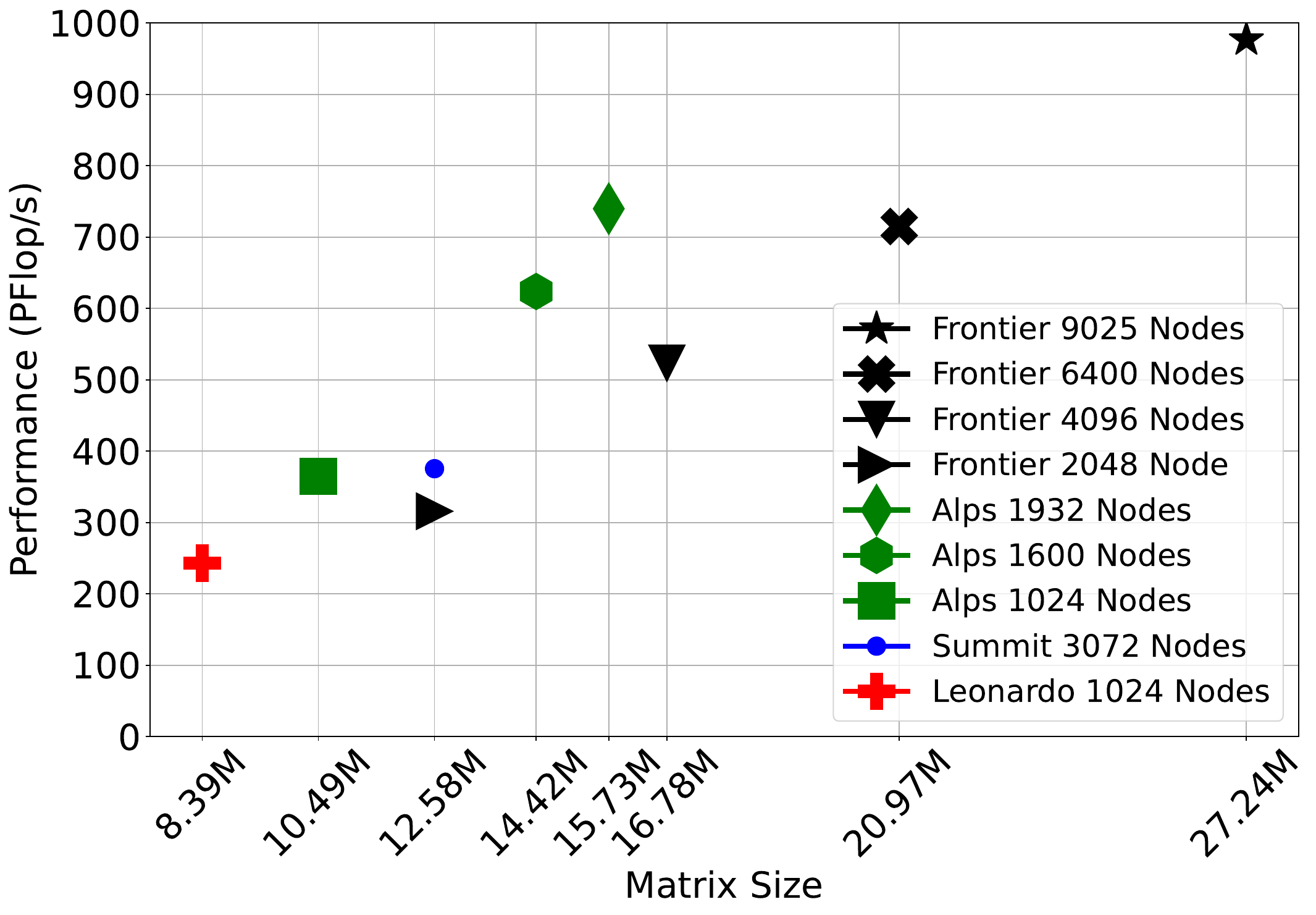}
\caption{\small Performance of largest runs on Summit, Leonardo, Alps, and Frontier; with additional
run-up points on Alps and Frontier, all using the DP/HP precision variant.}
\vspace{-5mm}
\label{fig:perf-large-all}
\end{figure}

Our work integrates state-of-the-art HPC to develop extreme-scale climate emulators, showcasing the potential of supercomputers to advance climate modeling at ultra-high resolution and enable 
significant savings in the petabytes of storage required for storing climate simulations. Our approach uses the SHT for modeling spatio-temporal climate data, 
accommodating climate data sourced from various spatial resolutions. We establish a sustainable approach to climate modeling on GPUs by shifting computation from communication-bound sparse double-precision kernels inherent in PDE-based models to low-precision tensor core kernels. 
As evidenced by IPCC assessment reports and our endorsement letters, the exascale climate emulator holds significant potential for the climate community, advancing climate research and policy making. Climate emulators have already found widespread application in assessing the sensitivity of various physical
parameters~\cite{Oakley2004Probabilistic}, performing model calibration \cite{chang2014}, analyzing emission
scenarios~\cite{Castruccio:2014}, studying the internal variability on climate model outputs~\cite{Hu2021Approximating}, investigating the persistence of climate anomalies and analyzing the temporal dynamics of climate systems~\cite{Branstator:2010}. Extending these applications, this work will enable climate scientists to apply ultra-high spatial resolution models to enhance understanding and predict climate change impacts.

This work also holds significant potential in advancing the development of machine learning (ML) and AI-driven methods for forecasting or prediction applications in climate science. For example, our framework can enhance hybrid approaches that combine physics with ML, thereby generating higher fidelity climate data~\cite{Yu:2024} and refine 3D neural networks recently proposed for weather forecasting~\cite{Kaifeng:2023}. Finally, we highlight the utility of our work for applications beyond climate science, given the flexible, scalable design of the emulator. For instance, our HPC innovations can be used to build an emulator for $N$-body simulations in cosmology for generating cosmological mass maps required for estimating the cosmological parameters~\cite{Perraudin:2021emulation}. 
In summary, we aim to drive the development of robust and multi-variate emulators for generating high-resolution spatio-temporal data that ride the waves of high performance architecture.

%% file: ack.tex
\small{
For computer time, this research used Shaheen III at the Supercomputing Laboratory of the King Abdullah University of Science and Technology (KAUST) in Thuwal, Saudi Arabia; Frontier and Summit at the Oak Ridge Leadership Computing Facility at the US DOE's Oak Ridge National Laboratory; Alps at CSCS in Lugano, Switzerland; and Leonardo at CINECA in Bologna, Italy. The authors are deeply grateful to the OLCF, CSCS, and CINECA for the discretionary allocations that allowed partial scaling runs after completing the science runs at KAUST and to NVIDIA for brokering the CSCS and CINECA connections. We also would like to thank Bilel Hadri from the KSL team at KAUST, as well as Thomas Herault, Aurelien Bouteiller, and Joseph Schuchart from the ICL team at the University of Tennessee, Knoxville (UTK), for their essential support in facilitating this work.
}